# Simulating Light Propagation through Biological Media Using Monte-Carlo Method

by

Maryam Ghahremani

Research Intern

July 2021

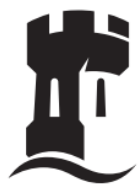

The University of Nottingham

# Abstract

Biological tissues are complex structures composed of many elements which make light-based tissue diagnostics challenging. Over the past decades, Monte Carlo technique has been used as a fundamental and versatile approach toward modeling photon-tissue interactions. This report first describes a MC simulation of steady-state light transport in an absorbing and diffusing multi-layered structure. Further, a parallel processing solution is implemented to reduce execution time and memory requirements. Then, the nonparametric phase function, which is a discretized version of the phase function, has been discussed where the integration of the phase function is a numerical process, instead of an analytical operation. Finally, to simulate more realistic structures of biological systems, simulations are modified to incorporate objects of various shapes (sphere, ellipsoid, or cylinder) with a refractive-index mismatched boundary. The output files mainly contain 2D reflection matrix, 2D transmission matrix, and 3D absorption matrix.

# 1. Introduction

Advanced computational methods enable accurate modeling of photon propagation behavior in diverse complex optical media and multilayer structures. Such numerical techniques have provided insight into light transmission dynamics ranging from integrated photonic components like coupled waveguides [1] to biological interfaces such as multilayer tissue [this tutorial], as well as emerging optical metamaterials [2], perovskite photovoltaic absorbers [3], and nanophotonic devices for quantum computing [4]. By leveraging computational electromagnetics and photonics principles, the photon distribution and interactions can be effectively simulated across these myriad systems to support novel applications.

Knowing the light fluence distribution or the amount of light scattered off the specimen is useful in many applications, such as photo-acoustic (PA) imaging [5], diffused optical tomography [6], Raman spectroscopy [7], and fluorescence imaging [8]. The light propagation models in biomedical optics predict light intensities, which are subsequently compared to measured light intensities on the tissue boundary. Based on the predicted and measured data, image reconstruction algorithms recover either the spatial distributions of intrinsic optical tissue properties or the concentrations of light emitting molecular probes inside tissue [9]. Note that the image reconstruction procedure requires a model for light propagation inside the target.

To model light interaction with biological tissue, the first step is typically to calculate the distribution of light within the tissue, given the optical properties for a given illumination. Most light propagation models are based or derived from the radiative transfer equation (RTE), which simplifies Maxwell's equations as it does not include non-linear properties of light such as interference, polarization and diffraction [10, 11]. But, which particular light propagation model is used depends on the optical wavelength of the light as well as the spatial size of the tissue domain interacting with the light.

The light-tissue interaction is governed by the optical tissue parameters, such as the *scattering*, $\mu_s$, and the *absorption*, $\mu_a$, coefficients, which are wavelength-dependent. The absorption coefficient, which can vary over several orders of magnitude, increases towards the visible wavelengths. Typical absorption parameters are in the range of 0.5-5 $cm^{-1}$ at wavelengths $\lambda < 625$ nm. In the red and near-infrared regions where $\lambda > 625$ nm the absorption coefficient varies between 0.01 and 0.5 $cm^{-1}$. On the other hand, the scattering coefficient varies only slightly as a function of the wavelength between 10 and 200 $cm^{-1}$[12]. Light scattering events in biological tissues are strongly forward-peaked and are well-described by the Henyey–Greenstein scattering kernel with the *mean scattering cosine* (also termed anisotropy factor) $g$, which varies typically between 0.5 and 0.95 depending on the tissue type. The *reduced scattering coefficient* is defined as $\mu'_s = (1-g)\mu_s$ incorporating both $\mu_s$ and $g$ to gain an equivalent value of scattering coefficient that would describe the overall photon transport within an isotropic tissue ($g = 0$) on a macroscopic scale. Typical reduced scattering coefficients are between 4 and 15 $cm^{-1}$ and are slightly wavelength-dependent. The *mean free path* (MFP) is the length $1/(\mu_s + \mu_a)$, whereas the *transport mean free path* (TMFP) is defined as the length $1/(\mu'_s + \mu_a)$, which plays an important role in diffusion theory. Refer to [13, 14] for a comprehensive review of optical tissue parameters.

Analytical solution for solving the RTE exists for simple cases and cannot be found for biological tissue with spatially non-uniform scattering and absorption properties and curved tissue geometries. Most light propagation models are based on the diffusion approximation to the RTE, when the condition $\mu_a \ll \mu'_s$ holds. While diffusion theory for photon transport is a fast and convenient way to model light propagation, it fails in the proximity of the light sources or boundaries and when absorption is strong compared to scattering; in other words, whenever conditions cause the gradient of fluence rate (or photon concentration) to not be simply linear but to have some curvature [15]. For more realistic media with complex multiple scattering effects such as cerebrospinal fluid (CSF) in the brain, lungs, and synovial fluid in the joints, numerical methods are required. Monte Carlo (MC) for light propagation in biological medium was first introduced by Wilson, but the implementation details of MC were discussed a decade later [16]. It was modified by many others for usability [17, 18].

MC refers to a technique first proposed by Metropolis and Ulam to simulate physical processes using a stochastic model [19].The MC simulations may be used for both diagnostic and therapeutic applications of lasers and other optical sources in medicine. For example, MC-simulated diffuse reflectance can be used to deduce optical properties of tissues, which may be used to differentiate cancerous tissue from normal tissue. MC-simulated optical energy deposition inside tissue may be used to compute light dosage for photodynamic therapy of disease. The advantage of MC simulation method lies in first providing accurate results, whatever the inter-fiber or source-detector distances, and second in providing the detailed spatial distribution of the fluence while taking into account a large number of parameters: the intensity spectrum of the excitation source, the locations and geometries of both illumination and light collection, the local coefficients of absorption and diffusion, the refractive indices and dioptres, etc.

As a stochastic solver to the RTE, the MC method consists in propagating a packet of elementary energy (photon or group of photons) step by step, with random sampling of the length and direction of the displacement steps, where the scattering angle is typically drawn from the inverse cumulative distribution of the scattering (or "phase") function. The heuristic Henyey-Greenstein (HG) phase function [20, 21] is most commonly used in the biomedical community since it offers an analytical inverse of the cumulative distribution function (CDF), which is convenient for sampling the scattering angles by a random number drawn from a uniform distribution. At the end of each photon step, the weight of the photon is reduced by absorption and the remaining non-absorbed weight is redirected according to the specified phase function. Once a new trajectory is identified, the photon is again moved a random distance. If the photon escapes from the tissue, the reflection or transmission of the photon is recorded. This process is repeated until the desired number of photons have been propagated.

In a MC simulation, the recorded quantities such as reflection, transmission, and absorption profiles will approach true values (for a tissue with the specified optical properties) as the number of photons propagated approaches infinity. MC, therefore, suffers from low computational efficiency because a large number of particle histories have to be simulated to achieve a desired statistical accuracy. Sequential MC simulations require extensive computation and long runtimes, easily taking up to several hours [22]. In recent years, studies on massively parallel MC algorithms have successfully reduced this computational cost down to seconds or minutes, due largely to the "embarrassingly parallelizable" nature of MC and the rapid adoption of low-cost many-core processors such as general-purpose graphics processing units (GPUs). Alerstam *et al.* [23] first reported a proof-of-concept using GPUs to accelerate MC in a homogeneous domain. In 2009, Fang and Boas [24] reported the first

GPU-accelerated MC algorithm to model light transport inside a three-dimensional (3-D) heterogeneous domain, and released an open-source tool—Monte Carlo eXtreme (MCX). For those who haven't a GPU and run the program on MATLAB platform, Parallel Computing Toolbox™ (PCT) can be used to speed up the simulations, the approach as we have used in this report.

The HG phase function is well known to underestimate large-angle backward scattering observed in turbid media such as tissue. More complex parametric phase functions, such as modified HG, two-term HG and Gegenbauer kernel, were thus proposed for improving the accuracy. They are, however, always an approximation of the real phase function. Therefore, nonparametric phase function, which is a discretized version of the phase function, has been studied and proposed by several researchers [25, 26]. This method is based on constructing a look-up table and the phase function is numerically sampled. Section 3 gives a detailed discussion in this regard.

MC Multilayer (MCML) model was a highly simplified model for many applications. Hence, MCML was further modified to incorporate objects of regular shapes, such as sphere and cylinder of matched (or un-matched) refractive index to mimic tumor and blood vessels. In section 4, we discuss this context and provide some results.

This report aims to revisit the concept of MC simulations, along with the discussions on the advancements in the simulation techniques. Steps involved in tracking of photons in tissue using MC are discussed in section 2. Further, a possible parallel solution for reducing execution time and memory requirements is investigated. Section 3 discusses implementations of the numerical phase function sampling for MC simulations and describes a sampling scheme utilizing linearly-spaced tabulated CDF values. In section 4, MC simulation of tissues with embedded objects is discussed. Final discussions are given in section 5.

# 2. The Steady-State Monte Carlo Propagation of Photons in a Tissue

The MC method uses random sampling from well-defined probability distributions to model the actions of a photon in turbid media and the transport of each photon is modelled as a random walk. The overall effect of the medium on the temporal and spatial distribution of photons as they travel through the medium can be found by running and tracking multiple photons, that is the sampling of all possible outcomes.

The propagation of photons in a MC simulation can be broken down in a few standard steps: propagation (hop), absorption (drop), scattering (spin), and the Fresnel phenomena of transmission and reflection at boundaries, where the refractive index changes. The details of the theory and operation of this model is discussed in what follows.

## 2.1 Tissue Representation

Tissue is represented as finite homogeneous layers where boundaries between layers are parallel to each other (Fig. 2.1). Photon packets are launched into the top layer, and propagate through the layers until the photon is terminated. A photon packet only leaves the tissue either at the top or bottom boundary. Each layer is infinitely wide, and is described by the following parameters: the thickness, the refractive index, the absorption coefficient $\mu_a$(cm$^{-1}$), the scattering coefficient $\mu_s$(cm$^{-1}$), and the anisotropy factor $g$. The refractive indices of the ambient medium above the tissue (e.g. air) and the ambient medium below the tissue (if existing) need to be given as well. Although the real tissue can never be infinitely wide, it can be so treated on the condition that it is much wider than the spatial extent of the photon distribution. The absorption coefficient $\mu_a$, is defined as the probability of photon absorption per unit infinitesimal path length, and the scattering coefficient $\mu_s$ is defined as the probability of photon scattering per unit infinitesimal path length. For the simplicity of notation, the total interaction coefficient $\mu_t$, which is the sum of the absorption coefficient $\mu_a$ and the scattering coefficient $\mu_s$, is sometimes used. Consequently, the interaction coefficient means the probability of photon interaction per unit infinitesimal path length. The anisotropy $g$ is the average of the cosine value of the deflection angle.

Three coordinate systems are used in the MC simulation at the same time. A Cartesian coordinate system is used to trace photon movements. The origin of the coordinate system is the photon incident point on the tissue surface; the $z$-axis is the normal of the surface pointing toward the inside of the tissue; and the $xy$-plane, is therefore on the tissue surface (Fig. 1).

For an infinitely narrow photon beam, since it is perpendicular to the tissue surface of a multilayered tissue structure, the problem has cylindrical symmetry. Therefore, we set up a cylindrical coordinate system to score internal photon absorption as a function of $r$ and $z$, where $r$ and $z$ are the radial and $z$ coordinates of the cylindrical coordinate system, respectively. The cylindrical coordinate system and Cartesian coordinate system share the origin and $z$-axis. The $r$ coordinate of the cylindrical coordinate system is also used for the diffuse reflectance and diffuse transmittance as a function of $r$ and $\alpha$, where $\alpha$ is the azimuthal angle.

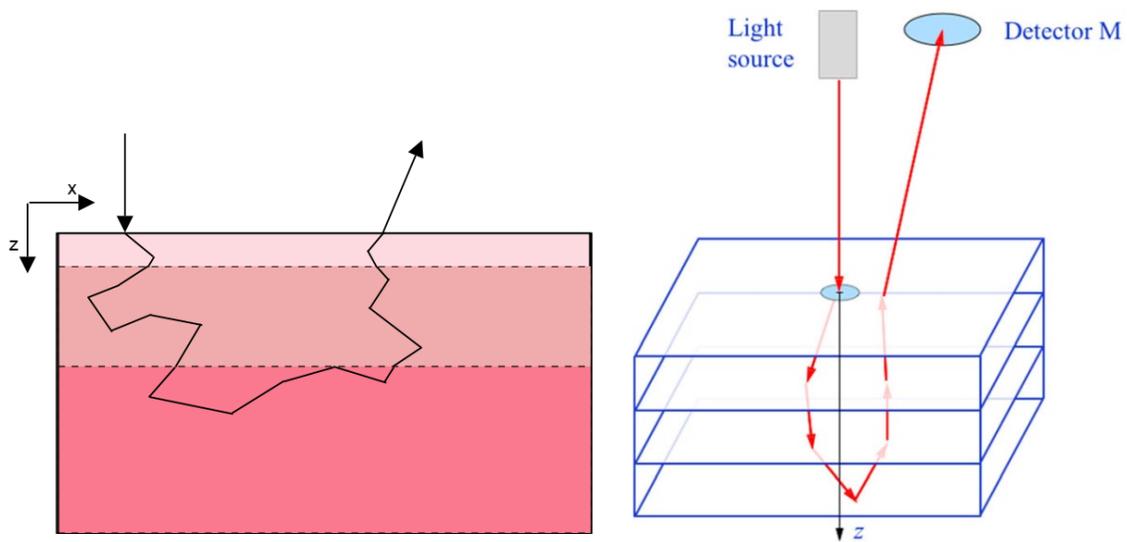

**Fig. 2.1.** Layered tissue representation, including a photon trace

## 2.2 Rules for Photon Propagation

**2.2.1. Photon Representation.** MC optical modeling does not treat the photon as a wave phenomenon. Although phase and polarization phenomena can be represented by MC methods, it does not play an important role in the focus of this simulation: energy transport.

One simple technique to improve the efficiency of a MC program is to propagate multiple individual photons (a packet) along each pathway. So, the method depicts each photon as a statistical packet of energy with a given amount of initial energy, its weight, $W$. Variance reduction techniques (implicit photon capture), such as weighting, are used to reduce the number of photons necessary to achieve the desired accuracy for a MC calculation [27]. Due to the probability of absorption, the photon packet loses some of its weight after each propagation step. Representing the photon as a packet increases the number of tissue interactions before the photon has no more energy left and loses its usefulness.

At any point in the simulation a photon packet has three properties associated with it: position, direction, and weight. The spatial position of a photon is specified using three Cartesian coordinates: $x$, $y$, and $z$. The photon packet direction of propagation is specified using direction cosines $(\mu_x, \mu_y, \mu_z)$ and weight is specified by $W$. The representation is shown in Fig. 2.2.

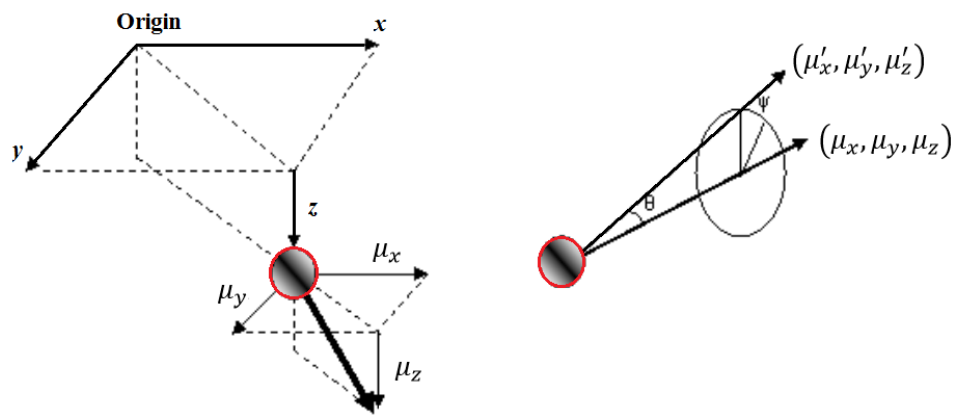

**Fig. 2.2.** Photon representation in Cartesian coordinates, and directional cosines.

**2.2.2. Photon Launch.** Photons are initially launched into the medium with an initial direction and location. Most of the sources apply a random sampling technique to distribute the photon launch locations equally over the source area. The direction is calculated using the properties defined for the source type, and a random sampling to distribute the directions equally. The MC simulation described in this report deals with the transport of an infinitely narrow photon beam, perpendicularly incident on a multi-layered tissue structure (Fig. 2.1). The responses to the infinitely narrow photon beam are called impulse responses.

After launch of the photon, if there is a mismatch between the refractive index of outside medium and tissue, specular reflection occurs. The amount of specular reflection is determined by Equation 2.1 where the refractive indices of the tissue and the outside medium are $n_1$ and $n_2$.

$$R_{sp} = \frac{(n_1 - n_2)^2}{(n_1 + n_2)^2} \tag{2.1}$$

If the first layer is glass, which is on top of a layer of medium whose refractive index is $n_3$, multiple reflections and transmissions on the two boundaries of the glass layer are considered. The specular reflectance is then computed by:

$$R_{sp} = r_1 + \frac{(1 - r_1)^2 r_2}{1 - r_1 r_2} \tag{2.2}$$

where $r_1$ and $r_2$ are the Fresnel reflectances on the two boundaries of the glass layer:

$$r_1 = \frac{(n_1 - n_2)^2}{(n_1 + n_2)^2} \tag{2.3}$$

$$r_2 = \frac{(n_3 - n_2)^2}{(n_3 + n_2)^2} \tag{2.4}$$

The weight of the photon packet is decremented by

$$W = 1 - R_{sp} \tag{2.5}$$

If we want to strictly distinguish the specular reflectance and the diffuse reflectance, we can keep track of the number of interactions experienced by a photon packet. When we score the reflectance, if the number of interactions is not zero, the reflectance is diffuse reflectance. Otherwise, it is specular reflectance. The transmittances can be distinguished similarly. To make a better understanding, Fig. 2.3 clearly illustrates the difference between a diffuse and specular reflectance.

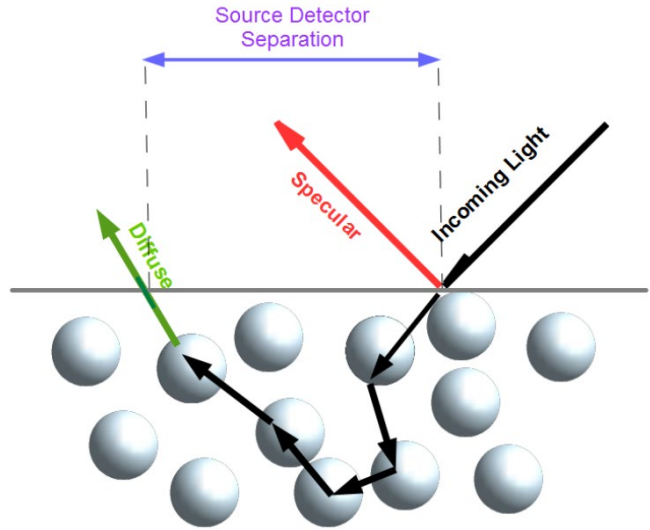

**Fig 2.3.** Schematic representation of specular and diffuse reflectance

**2.2.3. Basic Monte Carlo Sampling.** The MC simulation of light propagation in tissue requires random selection of photon step size, scattering angle and reflection or transmission at boundaries. This is accomplished by a random number [0, 1] assigned to the value of a random variable $x$, such as the photon step size. The relationship is established through the density function $p(x)$ and cumulative distribution function $F(x)$. Given $p(x)$, the value of the probability distribution function at a particular value $x_1$ of the random variable $x$ is

$$F(x_1) = \int_0^{x_1} p(x) dx \qquad (2.6)$$

The $F(x_1)$ is equated with a computer-generated pseudo-random number, $RND_1$, in the interval [0, 1]:

$$F(x_1) = RND_1$$

Then Eq. (2.6) is rearranged to solve for $x_1$ in terms of $RND_1$. The resulting expression allows a series of values $RND_1$ to specify a series of values $x_1$. The histogram of $x_1$ values will conform to the probability density function $p(x)$. Figure 2.4 illustrates the procedure. This process is used in the following by the random selection of photon step size and scattering angle.

**Fig. 2.4.** A random number generator [0, 1] selects a value $RND_1$ which is set equal to the cumulative distribution function $F(x)$ which then specifies the value $x_1$. The cross hatched area is equal to $RND_{1\,[28]}$.

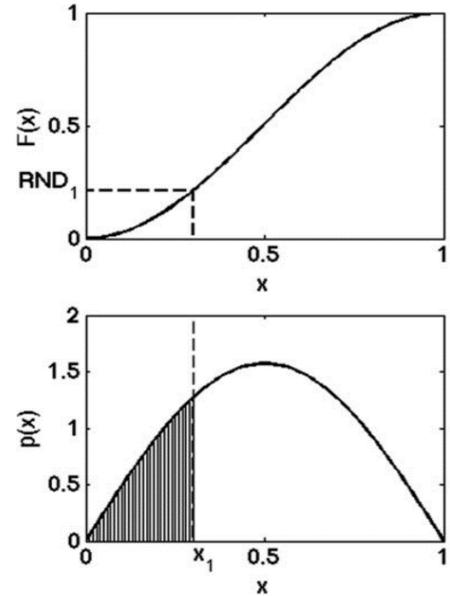

**2.2.4. Step Size of a Photon.** The distance a single photon (not a packet) travels in the tissue between interaction points is chosen randomly according to the optical properties of the tissue. A light beam propagating through a turbid media is subject to attenuation and scattering. The probability that a photon (or ray) scatters after travelling a distance $s$ follows an exponential probability distribution function defined by a mean-free path, $\mu_s^{-1}$, as

$$p(s) = \frac{e^{-\mu_s s}}{\mu_s} \qquad (2.7)$$

such that

$$\int_0^\infty p(s)\,ds = \int_0^\infty \frac{e^{-\mu_s s}}{\mu_s}\,ds = 1$$

where $\mu_s$ refers to the scattering coefficient of element the packet is located. Note that here the trajectory of the packet is not affected by the absorption coefficient. Instead, absorption coefficient along the path is used only to calculate a weight for the trajectory by Beer-Lambert law. This approach can be formally motivated by considering the solution of the radiative transfer equation for a ray-like initial condition (see e. g. 'microscopic Beer-Lambert law' in Ref. [29]) and noticing that the absorption coefficient enters in an exponential weighting factor for the intensity along the ray.

The integral of $p(s)$ evaluated at a particular $s_1$ is determined, which defines the probability distribution function as:

$$F(s_1) = \int_0^{s_1} p(s)ds \tag{2.8}$$

$$RND_1 = F(s_1) = \int_0^{s_1} \frac{e^{-\mu_s s}}{\mu_s}ds = 1 - e^{-\mu_s s_1} \tag{2.9}$$

and solve for $s_1$ as:

$$s_1 = \frac{-\ln(1 - RND_1)}{\mu_s} = \frac{-\ln(RND_1)}{\mu_s} \tag{2.10}$$

The terms $(1 - RND_1)$ and $(RND_1)$ are equal in a probabilistic sense because of the uniform distribution of the random number between $[0, 1]$.

**2.2.5. Hop.** The standard procedure for taking the step is described as the "Standard Hop" in Section 2.2.5.A But if the tissue has a front and/or rear boundary, then a second step to "Check boundaries" is taken, as described in Section 2.2.5.B, and if the photon is attempting to escape the tissue, a procedure is used to decide whether the photon escapes or is reflected back into the tissue.

### 2.2.5. A) Standard Hop (no boundary conditions)

The step size (scattering length) of the photon's step (or hop) must be determined. The step size is calculated:

$$\Delta s = \frac{-\ln(RND_1)}{\mu_s} \tag{2.11}$$

as was described in Section 2.2.4. Now the position of the photon is updated:

$$\begin{aligned} x &= x + \Delta s \mu_x \\ y &= y + \Delta s \mu_y \\ z &= z + \Delta s \mu_z \end{aligned} \tag{2.12}$$

### 2.2.5. B) Check Boundaries

As part of the Hop step, there is a side box labeled "check boundaries", which is used when there are front and surface boundaries in the problem. This boundary check is denoted by a side box to emphasize that it is part of the "Hop" step.

The photon moves freely through tissue until change in optical properties occurs. That point is a boundary between two different media (Fig. 2.5). This boundary can be between two media in the simulation area, or between the surrounding medium and a medium in the simulation area. For example, the photon packet may attempt to escape the tissue at the air/tissue interface. If this is the case, then the photon packet may either escape as observed reflectance (or transmittance if a rear boundary is also included) or be internally reflected by the boundary. There are different methods of dealing with this problem when the step size is large enough to hit the boundary. Let us present the approach we have used in the program MCML.

When a step size $\Delta s$ is large enough to cross a boundary between media, it is split up in two parts:

$$\Delta s = \Delta s_{to-boundary} + \Delta s_{remaining} \tag{2.13}$$

First, a foreshortened step size $\Delta s_{to-boundary}$ is computed as:

$$\Delta s_{to-boundary} = \begin{cases} (z - z_0)/\mu_z & \text{if } \mu_z < 0 \\ (z - z_1)/\mu_z & \text{if } \mu_z > 0 \end{cases} \tag{2.14}$$

where $z_0$ and $z_1$ are the $z$ coordinates of the upper and lower boundaries of the current layer. The foreshortened step size $\Delta s_{to-boundary}$ is the distance between the current photon location and the boundary in the direction of the photon propagation. Since the photon direction is parallel with the boundary when $\mu_z$ is zero, the photon will not hit the boundary. Therefore, Eq. 2.14 does not include the case when $\mu_z$ is zero. We move the photon packet $\Delta s_{to-boundary}$ to the boundary with a flight free of interactions with the tissue and the photon takes the partial step $\Delta s_{to-boundary}$ as

$$\begin{aligned} x &= x + \mu_x \Delta s_{to-boundary} \\ y &= y + \mu_y \Delta s_{to-boundary} \\ z &= z + \mu_z \Delta s_{to-boundary} \end{aligned} \tag{2.15}$$

Second, we compute the probability of a photon packet being internally reflected, which depends on the angle of incidence, $\alpha_i$, onto the boundary, where $\alpha_i = 0$ means orthogonal incidence. The value of $\alpha_i$ is calculated from directional cosine $\mu_z$:

$$\alpha_i = \cos^{-1}\left(\left|\mu_z\right|\right) \tag{2.16}$$

The critical angle, $\alpha_c$, for the boundary determines if the photon is completely internally reflected. The critical angle is calculated by:

$$\alpha_c = \sin^{-1}\left(n_t/n_i\right) \tag{2.17}$$

Where $n_t$, $n_i$ are the refractive indices of the transmission medium and incidence medium respectively. If a fraction of weight is transmitted to the new medium, the new angle, $\alpha_t$, is determined by Snell's law

$$n_t \sin\alpha_t = n_i \sin\alpha_i \tag{2.18}$$

The reflected part of the photon, $R_i$, is determined by Fresnels's laws:

$$R_i = \begin{cases} \dfrac{\left(n_i - n_t\right)^2}{\left(n_i + n_t\right)^2} & \text{if } \alpha_i = 0 \\[2em] \dfrac{\left(\sin\alpha_i \cos\alpha_t - \cos\alpha_i \sin\alpha_t\right)^2}{2} \\[1em] \times\dfrac{\left[\left(\cos\alpha_i \cos\alpha_t + \sin\alpha_i \sin\alpha_t\right)^2 + \left(\cos\alpha_i \cos\alpha_t - \sin\alpha_i \sin\alpha_t\right)^2\right]}{\left[\left(\sin\alpha_i \cos\alpha_t + \cos\alpha_i \sin\alpha_t\right)^2 \left(\cos\alpha_i \cos\alpha_t + \sin\alpha_i \sin\alpha_t\right)^2\right]} & \text{if } 0 < \alpha_i < \alpha_c \\[2em] 1 & \text{if } \alpha_i > \alpha_c \end{cases} \tag{2.19}$$

where in the second relation it is assumed that the simulated photon's polarization is evenly distributed between $s$ and $p$ states.

The next step of the photon is determined by the type of boundary. It is positioned at either an outer boundary, or an internal boundary. The distinction between these is below:

- When the photon is positioned at the 'outer' boundary, a part of the photon, $W_t$, is transferred into the surrounding medium:

$$W_t = \left(1 - R\left(\alpha_i\right)\right) * W \tag{2.20}$$

This weight must be incremented at the particular grid element as the reflectance or transmittance. Therefore, the reflectance, $R_d(r, \alpha_t)$, or transmittance, $T_t(r, \alpha_t)$, is updated by the amount of escaped photon weight, $W_t$:

$$R_d(r, \alpha_t) = R_d(r, \alpha_t) + W_t \quad \text{if} \quad z = 0$$

$$T_t(r, \alpha_t) = T_t(r, \alpha_t) + W_t \quad \text{if} \quad z = \text{the bottom of the tissue.}$$

(2.21)

The new weight of the photon, $W'$, is updated to the reflected part of the photon's weight, $W$, as:

$$W' = R(\alpha_i) * W$$

(2.22)

The new direction is chosen by reversing the directional cosine $\mu_z$.

- The new path and weight of a photon at an 'inner' boundary is determined by a random variable $\xi$. If $\xi \leq R(\alpha_i)$, the photon is reflected internally and its directional cosines $(\mu_x, \mu_y, \mu_z)$ must be updated by reversing the $z$ component, i.e., $(\mu_x, \mu_y, \mu_z) \rightarrow (\mu_x, \mu_y, -\mu_z)$ and the new weight is determined by Eq. 2.22.

**Layer Transition:**

Otherwise, the photon is transmitted to the other layer of tissue, and it has to continue propagation instead of being terminated. The new weight after transmission is determined by:

$$W' = (1 - R(\alpha_i)) * W$$

(2.23)

The new directional cosines, $(\mu'_x, \mu'_y, \mu'_z)$ are determined by

$$\mu'_x = \mu_x \frac{\sin \alpha_t}{\sqrt{1 - \mu_z^2}} = \mu_x \frac{n_i}{n_t}$$

$$\mu'_y = \mu_y \frac{\sin \alpha_t}{\sqrt{1 - \mu_z^2}} = \mu_y \frac{n_i}{n_t}$$

(2.24)

$$\mu'_z = \cos \alpha_t \frac{z_{boundarry} - z}{|z_{boundarry} - z|}$$

and, the remaining step size has to be converted for the new tissue according to its optical properties:

$$\Delta s'_{remaining} = \frac{\mu_t}{\mu'_t} \Delta s_{remaining} \qquad (2.25)$$

where $\mu'_t$ is the value of $\mu_t$ in the new tissue layer. The current step size $\Delta s' = \Delta s_{remaining}$ is again checked for another boundary or interface crossing. The above process is repeated until the step size is small enough to fit in one layer of tissue. At the end of this small step, the absorption and scattering are processed correspondingly.

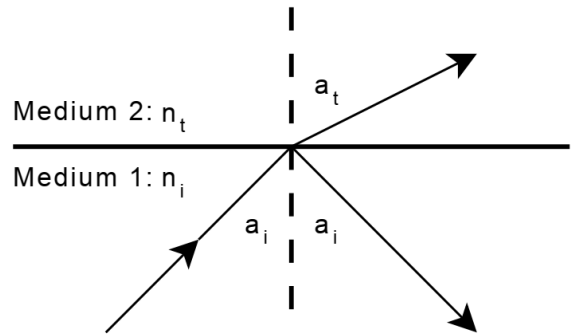

**Fig 2.5.** Boundary conditions to represent Snell's law

**2.2.6. Photon Absorption.**  The technique of implicit capture (variance reduction) assigns a weight to each photon as it enters tissue. After each propagation step, the photon packet is split into two parts—a fraction is absorbed and the rest is scattered. The fraction of the packet that is absorbed is

$$\Delta W = W_0 e^{-\mu_a \Delta s} \qquad (2.26)$$

Consequently, the new photon weight ($W$) is given by $W' = W_0 - \Delta W$ which represents the fraction of the packet that is scattered on this step.

An absorption event requires that both the location and the amount of light absorbed be recorded. The absorbed fraction is placed in the bin that encloses the current photon position. For example, the appropriate element of the absorption matrix is incremented by $\Delta W$. This process is summarized by the following calculation steps in Cartesian coordinate:

$$i_x = \frac{N_x}{2} + round\left(x/dx\right) + 1 \qquad (2.27.a)$$

$$i_y = \frac{N_y}{2} + round\left(y/dy\right) + 1 \qquad (2.27.b)$$

$$i_z = round\left(z/dz\right) + 1 \qquad (2.27.c)$$

$$A[i_x, i_y, i_z] = A[i_x, i_y, i_z] + \Delta W \qquad (2.27.d)$$

where $dx$, $dy$, $dz$ are the size of each bin along $x$, $y$, and $z$ axes, respectively, and $N_x$ and $N_y$ are the number of bins along $x$ and $y$ axes. In a Cylindrical coordinate system, we have

$$\alpha = \begin{cases} \cos^{-1}\left(x/r\right) & y \geq 0 \\ 2\pi - \cos^{-1}\left(x/r\right) & y \leq 0 \end{cases} \Rightarrow i_\alpha = \begin{cases} 1 & \text{if } \ r = 0 \\ round\left(\alpha/d\alpha\right) + 1 \end{cases} \qquad (2.28.a)$$

$$i_r = round\left(r/dr\right) + 1 \qquad (2.28.b)$$

$$i_z = round\left(z/dz\right) + 1 \qquad (2.28.c)$$

$$A[i_r, i_\alpha, i_z] = A[i_r, i_\alpha, i_z] + \Delta W \qquad (2.28.d)$$

where $dr$, $d\alpha$, $dz$ are the size of each bin along $r$, $\varphi$, and $z$ axes, respectively, and $N_r$, $N_\alpha$, and $N_z$ are the number of bins along each axis. Note that the number of bins in the absorption matrix is determined by the spatial resolution desired. Increasing the number of entries increases the spatial resolution, but also increases the absorption uncertainty in each element (because fewer absorption events will take place in each element and the error is inversely proportional to the square root of the number of absorption events). The fluence rate is obtained by dividing the final value of each matrix element by (1) the equivalent spatial volume of the element, (2) the absorption coefficient, (3) the total number of photons propagated, and (4) the initial weight of each photon.

**2.2.7. Photon Scattering.** Once the photon packet has been moved, and its weight decremented, only when $\mu_a$ is not 0, the photon packet is given a new direction as the result of a scatter event.

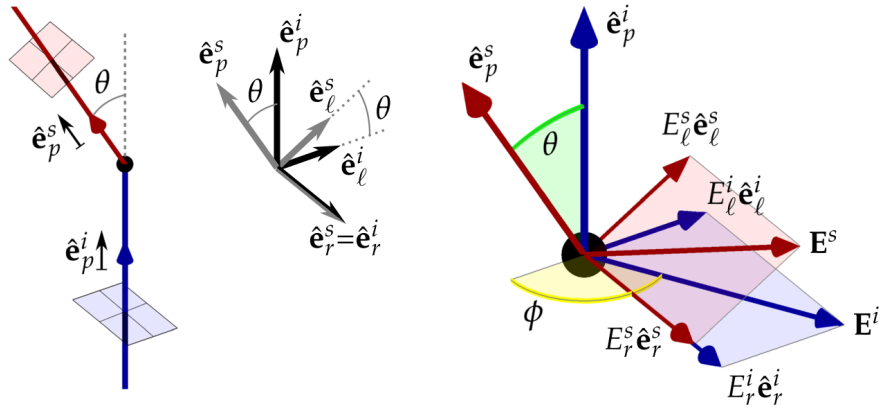

**Fig. 2.6** Geometry of a scattering event. Left diagram: incident ray vector $\hat{e}_p^i$, and scattered ray vector $\hat{e}_s^i$, define the scattering plane common to both vectors and $\theta$ is the scattering angle. Planes for defining polarization states are represented by the red and blue squares. Centre diagram: the orthonormal co-ordinate systems for $\hat{e}_p^i$, and $\hat{e}_s^i$; the $r$ direction, common to both systems, is orthogonal to the scattering plane; the propagation direction $p$ and the orthogonal $l$ direction for the scattered photon are rotated by $\theta$ in the scattering plane with respect to the incident ray. Right diagram: the electric field $\mathbf{E}_i$ of the incident ray (blue) and $\mathbf{E}_s$ for the scattered ray (red) are decomposed into the respective $r$ and $l$ directions, with respect to the azimuthal angle $\phi$; as for the left diagram, the polarization planes are shaded red and blue [30].

The nature of a scattering event is represented in Fig. 2.6. In the figure, the incident ray propagates in the direction of the unit vector $\hat{e}_p^i$ represented by the blue arrow in the left-most diagram, and is scattered at the location represented by the black dot towards the new scattered direction $\hat{e}_p^s$ represented by the red arrow. These vectors determine the scattering plane (drawn in green in the right-most diagram), defined by the scattering angle $\theta$. The plane of polarization of each ray (represented by the blue and red planes, where the electric field resides, perpendicular to the propagation directions) is defined by two unit vectors in the directions parallel and perpendicular to the scattering plane, and therefore $\left(\hat{e}_p^i,\ \hat{e}_r^i,\ \hat{e}_l^i\right)$ and $\left(\hat{e}_p^s,\ \hat{e}_r^s,\ \hat{e}_l^s\right)$ are the orthonormal vector basis oriented with the incident and scattered rays, respectively, where the propagation, perpendicular and parallel directions are denoted by sub-indices $p$, $r$, and $l$, respectively. The electric field of each ray (confined in the polarizations plane) can be decomposed into the parallel and perpendicular components, $\vec{E}^i = E_l^i \hat{e}_l^i + E_r^i \hat{e}_r^i$ and $\vec{E}^s = E_l^s \hat{e}_l^s + E_r^s \hat{e}_r^s$, where the mutual phases of the complex electric fields completely define arbitrary states of polarization.

There will be a deflection angle, $\theta \in [0, \pi)$, and an azimuthal angle, $\phi \in [0, 2\pi)$ to be sampled statistically. Note that the probability density functions for $p(\phi)$ and $p(\theta)$ are considered independently. The probability distribution for the cosine of the deflection angle, $\cos\theta$, is described by the scattering function that Henyey and Greenstein [20, 21] originally proposed for galactic scattering:

$$p(\cos\theta) = \frac{1}{2} \frac{1 - g^2}{\left(1 + g^2 - 2g\cos\theta\right)^{3/2}} \tag{2.29}$$

which has the properties that

$$\int_0^\pi p(\theta) 2\pi \sin\theta \, d\theta = 1 \tag{2.30}$$

$$\int_0^\pi p(\theta) \cos\theta \, 2\pi \sin\theta \, d\theta = g \tag{2.31}$$

where the last equation is the definition of $g$. Hence, the HG function is an identity with respect to the definition of $g$ between –1 and 1. If you choose a value $g$ to define $p(\theta)$ using Eq. (2.29), the definition of $g$ in Eq. (2.31) will yield exactly $g$. A value of $g = 0$ indicates isotropic scattering and a value near 1 indicates very forward directed scattering. Jacques *et al.* [31] determined experimentally that the Henyey-Greenstein function described single scattering in tissue very well. Values of $g$ range between 0.3 and 0.98 for tissues, but quite often $g$ is ~0.9 in the visible spectrum. Fig. 2.7 illustrates polar plots of HG phase function for different values of $g$.

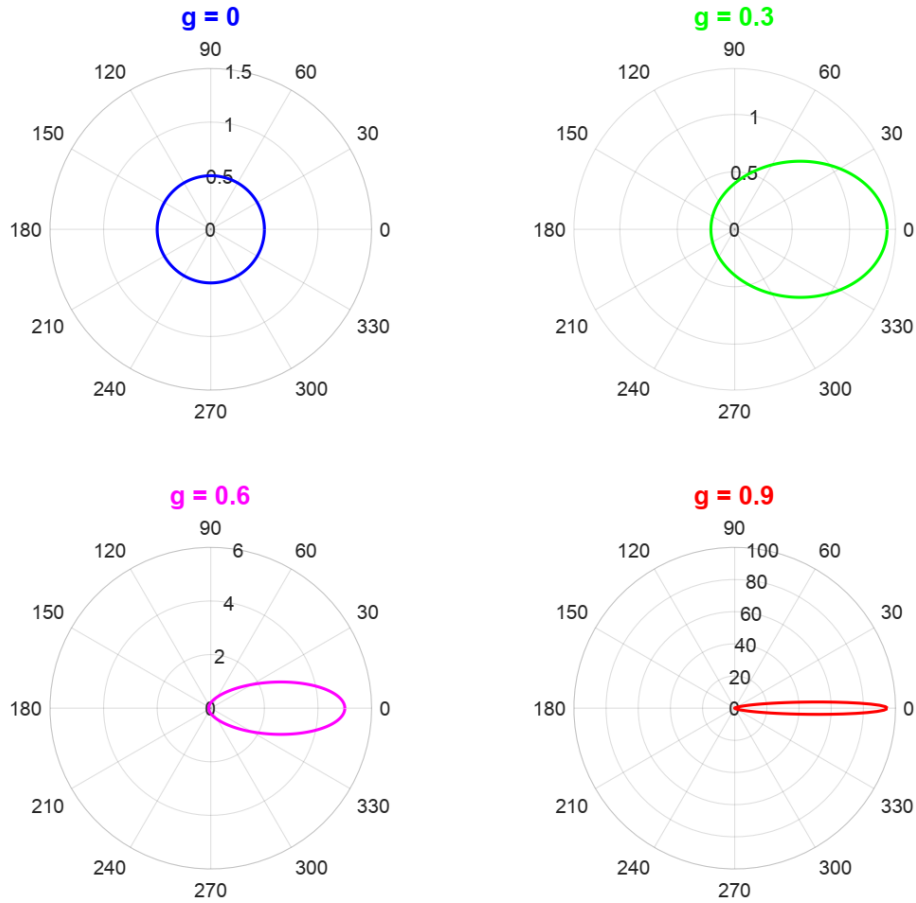

**Fig. 2.7** Polar plots of HG phase function for different values of $g$.

Applying Eq. (2.8), the choice for $\cos\theta$ can be expressed as a function of the random number, $\xi$ :

$$\cos\theta = \begin{cases} \dfrac{1}{2g}\left\{1 + g^2 - \left[\dfrac{1-g^2}{1-g+2g\,\xi}\right]^2\right\} & \text{if } g \neq 0 \\[4mm] 2\xi - 1 & \text{if g=0} \end{cases} \tag{2.32}$$

Next, the azimuthal angle, $\varphi$, which is uniformly distributed over the interval 0 to $2\pi$, is sampled:

$$\phi = 2\pi\,\xi \tag{2.33}$$

Using the calculated $\varphi$ and $\theta$, the new directional cosines are calculated by the following equations:

$$\mu'_x = \frac{\sin\theta\left(\mu_x\mu_z\cos\phi - \mu_y\sin\phi\right)}{\sqrt{1-\mu_x^{\,2}}} + \mu_x\cos\theta$$

$$\mu'_y = \frac{\sin\theta\left(\mu_y\mu_z\cos\phi - \mu_x\sin\phi\right)}{\sqrt{1-\mu_y^{\,2}}} + \mu_y\cos\theta \qquad (2.34)$$

$$\mu'_z = -\sin\theta\cos\phi\sqrt{1-\mu_z^{\,2}} + \mu_z\cos\theta$$

where $\left(\mu_x,\mu_y,\mu_z\right) = \left(\hat{e}_x^{\,i},\hat{e}_y^{\,i},\hat{e}_z^{\,i}\right)$ and $\left(\mu'_x,\mu'_y,\mu'_z\right) = \left(\hat{e}_x^{\,s},\hat{e}_y^{\,s},\hat{e}_z^{\,s}\right)$. If the trajectory is extremely close to alignment with the $z$-axis, *i.e.* nearly $\left(\mu_x,\mu_y,\mu_z\right) = \left(0,0,\pm 1\right)$ then the following formulas should be used:

$$\mu'_x = \sin\theta\cos\phi$$

$$\mu'_y = \sin\theta\sin\phi \qquad (2.35)$$

$$\mu'_z = \frac{\mu_z}{|\mu_z|}\cos\theta$$

Finally, the current photon direction is updated as: $\mu_x = \mu'_x$, $\mu_y = \mu'_y$, $\mu_z = \mu'_z$. The photon is now oriented along a new trajectory, and ready to take a new step $\Delta s$.

There are alternative scattering functions. Mie theory is an important scattering phase function to consider. Perhaps an experiment has yielded a particular scattering function, and one wishes to run a simulation using this function. This report will not discuss these alternatives, but as long as the criteria of Eq. (2.30) is followed, most any scattering function for the deflection angle $\theta$ can be used. Sometimes the scattering function does not lend itself to a solution of $\cos\theta$ in terms of a random number, as in Eq. (2.32). Also, sometimes one wishes to consider an azimuthal scattering angle $\phi$ that depends on the deflection angle $\theta$, as in Mie scattering of polarized light. In such cases, the "rejection method" or "tabulated method" can be a useful approach.

**2.2.8. Photon Termination.** After a photon packet is launched, it can be terminated naturally by reflection or transmission out of the tissue. For a photon packet that is still propagating inside the tissue, if the photon weight, $W$, has been sufficiently decremented after many steps of interaction such that it falls below a threshold value (e.g., $W_{th} = 0.0001$), then further propagation of the photon yields little information unless you are interested in the very late stage of the photon propagation. However, proper termination must be executed to ensure conservation of energy (or number of photons) without skewing the distribution of photon deposition. A technique called 'Russian roulette' is used to terminate the photon packet when $W \leq W_{th}$. The roulette technique gives the photon packet one chance in $m$ (e.g., $m = 10$) of surviving with a weight of mW. If the photon packet does not survive the roulette, the photon weight is reduced to zero and the photon is terminated.

$$W = \begin{cases} mW & \text{if} \quad \xi \leq 1/m \\ 0 & \text{if} \quad \xi > 1/m \end{cases} \tag{2.36}$$

where $\xi$ is the uniformly distributed pseudo-random number. This method conserves energy yet terminates photons in an unbiased manner. The combination of photon roulette and splitting that is contrary to roulette, may be properly used to reduce variance.

## 2.3 Flowchart of photon tracing

Fig. 2.8 indicates the basic flowchart for the photon tracing part of the MC calculation as described in section 2.2. Many boxes in the flowchart are direct implementations of the discussions in section 2.2. This chart also includes the situation where the photon packet is in a glass, in which absorption and scattering do not exist. In this case, the photon packet will be moved to the boundary of the glass layer in the current photon direction. Then, we have to determine statistically whether the photon packet will cross the boundary or be reflected according to the Fresnel's formulas.

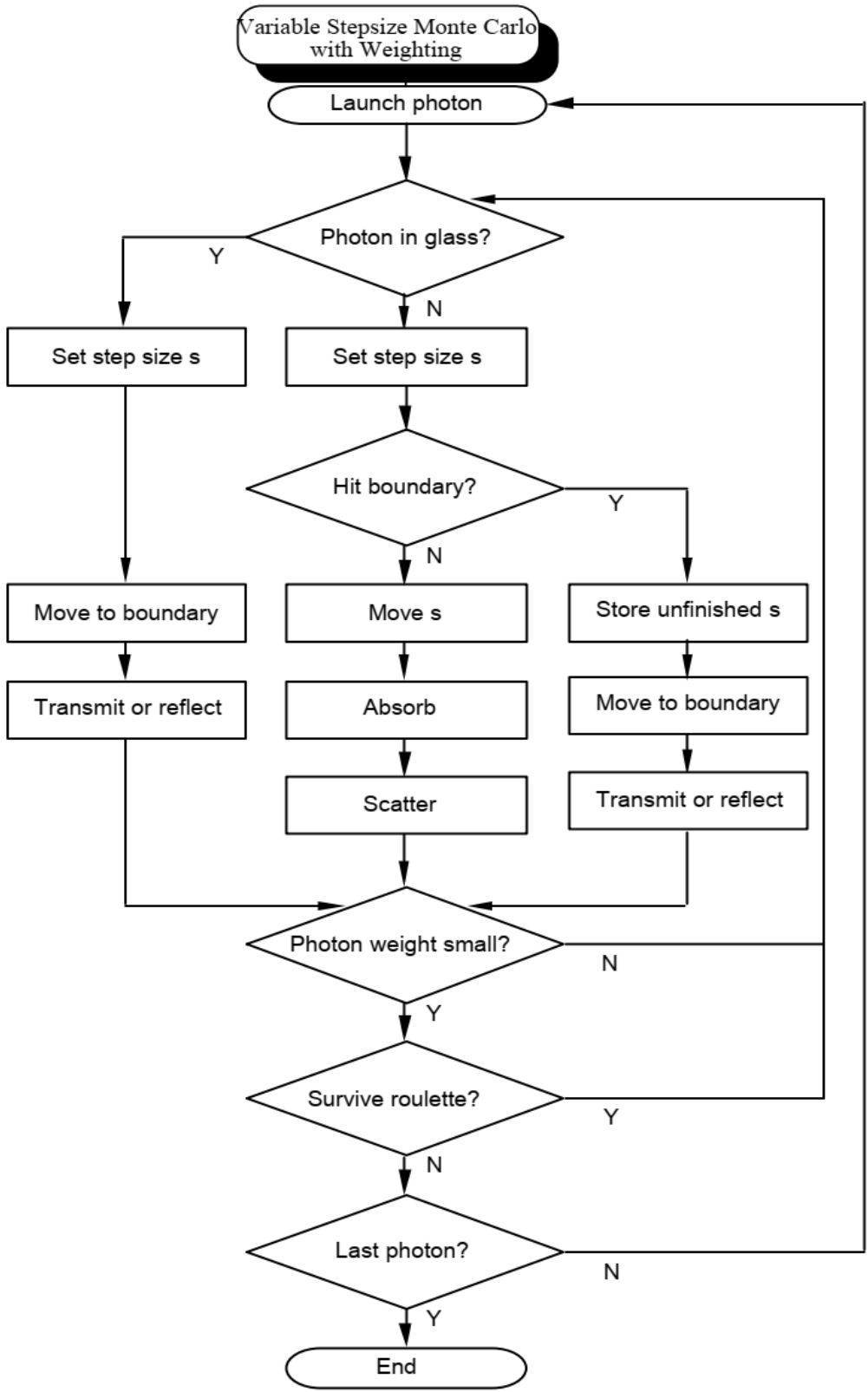

**Fig. 2.8.** A flowchart depicting the MC optical modeling of tissue.

## 2.4 Scored Physical Quantities

As we mentioned earlier, we record the photon reflectance, transmittance, and absorption during the MC simulation. To score these physical quantities, we need to set up grid systems. For scoring photon absorption, a two-dimensional homogeneous grid system is set up in the $r$ and $z$ directions. The grid separations are $dr$ and $dz$ in the $r$, and $z$ directions, respectively. The total numbers of grid elements in the $r$ and $z$ directions are $N_r$, and $N_z$, respectively. For scoring diffuse reflectance and transmittance, a two-dimensional homogeneous grid system is set up in the $r$ and $\alpha$ directions. This grid system can share the $r$ direction with the grid system for photon absorption; therefore, we need to set up one extra one-dimensional grid system for the diffuse reflectance and transmittance in the $\alpha$ direction. The total number of grid elements is $N_\alpha$.

Note that the last cells in $z$ and $r$ directions require special attention. Because photons can propagate beyond the grid system, when the photon weight is recorded into the diffuse reflectance or transmittance array, or absorption array, the physical location may not fit into the grid system. In this case, the last cell in the direction of the overflow is used to collect the photon weight. Therefore, the last cell in the $z$ and $r$ directions do not give the real value at the corresponding locations.

**2.4.1. Reflectance and transmittance**. When a photon packet is launched, the specular reflectance is computed immediately (Eq. (2.1)). The photon weight after the specular reflection is transmitted into the tissue. During the simulation, some photon packets may exit the media and their weights are accordingly scored into the diffuse reflectance array or the transmittance array depending on where the photon packet exits. After tracing multiple photon packets ($N$), we have two scored arrays $R_d(r, \alpha)$ and $T_t(r, \alpha)$ for diffuse reflectance and transmittance, respectively. They are internally represented by $R_{d-r\alpha}(i_r, i_\alpha)$ and $T_{t-r\alpha}(i_r, i_\alpha)$, respectively, in the program. The coordinates of the center of a grid element are computed by:

$$r = (i_r + 0.5)dr \quad [\text{cm}] \tag{2.37.$a$}$$

$$\alpha = (i_\alpha + 0.5)d\alpha \quad [\text{rad}] \tag{2.37.$b$}$$

where $i_r$ and $i_\alpha$ are the indices for $r$ and $\alpha$ (See Eqs. (2.28. a-d)). The raw data give the total photon weight in each grid element in the two-dimensional grid system. To get the total photon weight in the grid elements in each direction of the two-dimensional grid system, we sum the 2D arrays in the other dimension:

$$R_{d-r}[i_r] = \sum_{i_\alpha=0}^{N_\alpha-1} R_{d-r\alpha}[i_r, i_\alpha] \qquad (2.38.\,a)$$

$$R_{d-\alpha}[i_\alpha] = \sum_{i_r=0}^{N_r-1} R_{d-r\alpha}[i_r, i_\alpha] \qquad (2.38.\,b)$$

$$T_{t-r}[i_r] = \sum_{i_\alpha=0}^{N_\alpha-1} T_{t-r\alpha}[i_r, i_\alpha] \qquad (2.38.\,c)$$

$$T_{t-\alpha}[i_\alpha] = \sum_{i_r=0}^{N_r-1} T_{t-r\alpha}[i_r, i_\alpha] \qquad (2.38.\,d)$$

To get the total diffuse reflectance and transmittance, we sum the 1D arrays again:

$$R_d = \sum_{i_r=0}^{N_r-1} R_{d-r}[i_r] \qquad (2.39.\,a)$$

$$T_t = \sum_{i_r=0}^{N_r-1} T_{t-r}[i_r] \qquad (2.39.\,b)$$

All these arrays give the total photon weight per grid element, based on $N$ initial photon packets with weight unity.

**2.4.2. Internal photon distribution**. During the simulation, the absorbed photon weight is scored into the absorption array $A(r,z)$. $A(r,z)$ is internally represented by a 2D array $A_{rz}(i_r, i_z)$ where $i_r$ and $i_z$ are the indices for grid elements in $r$ and $z$ directions. The coordinates of the center of a grid element can be computed by Eq. (2.37) and the following:

$$z = (i_z + 0.5)dz \qquad (2.40)$$

The raw data $A_{rz}(i_r, i_z)$ give the total photon weight in each grid element in the two-dimensional grid system. To get the total photon weight in each grid element in the z direction, we sum the 2D array in the $r$ direction:

$$A_z[i_z] = \sum_{i_r=0}^{N_r-1} A_{rz}[i_r, i_z] \qquad (2.41)$$

The total photon weight absorbed in each layer $A_l$[layer] and the total photon weight absorbed in the tissue $A$ can be computed from $A_z[i_z]$:

$$A_l[\text{layer}] = \sum_{i_z \text{ in layer}}^{N_z-1} A_z[i_z] \qquad (2.42)$$

$$A = \sum_{i_z=0}^{N_z-1} A_z[i_z] \qquad (2.43)$$

where the summation range "$i_z$ in layer" includes all $i_z$'s that lead to a $z$ coordinate in the layer.

All the above arrays give the total photon weight per grid element [photon weight/bin] as $A_z(i_r, i_z)$ or perhaps $A_z(i_x, i_y, i_z)$, based on $N_{\text{photons}}$ initial photon packets with weight unity. Either way, the key parameter is the volume $V$[cm³] associated with each bin. For $A_z(i_r, i_z)$, the volumes vary with the value of $i_r$,

$$V = 2\pi(i_r + 0.5)dr^2 dz \qquad (2.44)$$

and for $A_z(i_x, i_y, i_z)$, the volumes are all equal,

$$V = dxdydz \qquad (2.44)$$

Then the values $A$ [photon weight/bin] are normalized by the appropriate $V$ and by the value $N_{\text{photons}}$ to yield the absorbed fraction, $A$ [1/cm³], for each pixel:

$$A(i_r, i_z) = \frac{A(i_r, i_z)}{V(i_r, i_z)N_{photons}} \qquad (2.45)$$

The light fluxes that have escaped at the front and rear surface boundaries are similarly normalized, but in this case the surface area $S$ rather than the volume $V$ is used. The value of $S[i_r]$ is $2\pi(i_r + 0.5)dr^2$ which is the area of annular ring. The array of escaping photons, $R_{d-r}[i_r]$ [photon weight/bin], is converted to the fractional escaping flux density, $R_{d-r}[i_r][1/cm^2]$, by the expression:

$$R_{d-r} = \frac{R_{d-r}}{S * N_{photons}} \qquad (2.46)$$

## 2.5 Sample Computation

Some computation results are described in this section as examples, and some of them are compared with the MC simulation results from other investigators to verify the program[32].

**2.5.1 Total diffuse reflectance and total transmittance.** We computed the total diffuse reflectance $R_d$ and total transmittance $T_t$ (including un-scattered transmittance) of a structure containing three slab layers of turbid media with the following optical properties: relative refractive indices $n_1 = 1.37$, $n_2 = 1.37$, $n_3 = 1.37$ (refractive index matched boundaries), absorption coefficients $\mu_{a1} = 1/\text{cm}$, $\mu_{a2} = 1/\text{cm}$, $\mu_{a3} = 2/\text{cm}$, scattering coefficients $\mu_{s1} = 100/\text{cm}$, $\mu_{s2} = 10/\text{cm}$, $\mu_{s3} = 10/\text{cm}$, anisotropy factors $g_1 = 0.9$, $g_2 = 0$, $g_3 = 0.7$, and thicknesses of $D_1 = 0.1$, $D_2 = 0.1$, $D_3 = 0.2$ [cm] as illustrated in Fig. 2.9. The refractive indices of the top and bottom ambient media are both set to 1.0. The MC simulations of $N_{photons} = 10^6$ packets each are completed which are delivered as a pencil beam of collimated light at the origin and from the top surface: ($x = 0$, $y = 0$, $z = 0$), ($\mu_x = 0$, $\mu_y = 0$, $\mu_z = 1$). Since the infinitely narrow photon beam is perpendicular to the surface of a multilayered tissue, the problem has cylindrical symmetry. Therefore, it is assumed that we are working in cylindrically symmetric Cartesian coordinates. This means the MC simulation will propagate the photon in 3D using $x$, $y$ and $z$ coordinates, but we can report out our results only as a function of $r$ and $z$. We assume the beam of light has a radius $r_0 = 0.5\text{cm}$. The number of grid elements along $r$ and $z$ axis is $N_r = N_z = 100$, thus we have $dr = r_0/N_r = 0.5/100$.

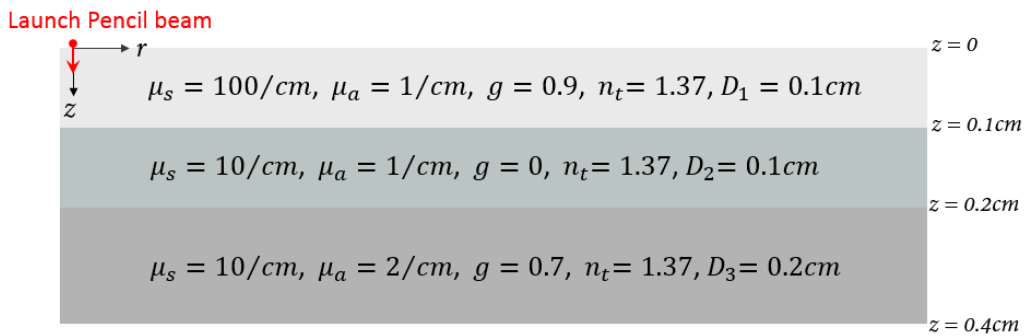

**Fig. 2.9.** MC simulated structure containing three slab layers of turbid media. Optical properties of each layer can be found in the figure.

Comparing the results obtained during simulation against known data from literature gives a good indication of our simulation correctness. Here, we compared our MC simulation results of multi-layered tissue with the results of an independently written MC simulation. The comparison includes the transmittance versus radius $r$ detected at the bottom surface ($z = z_{max}$), $T_t(r)$, where the radius $r$ is the distance between the photon incident point and the observation point, and the diffuse reflectance versus radius $r$ detected at the top surface ($z = 0$), $R_d(r)$ (See Eqs. 2.38(a) and (c)). All processing of output files and visualization is done using MATLAB platform. The normal HG phase function is used for sampling the deflection angle $\theta$. Here, the exiting angles of the reflected or transmitted photons, $\alpha$, are not resolved, therefore the number of grid elements in the angle $\alpha$ direction is set equal to 1. The obtained transmittance and diffuse reflectance from the two simulations are given in Fig. 2.10 (a) and (b). It can be seen that the comparisons have shown agreement between the two independently simulated results.

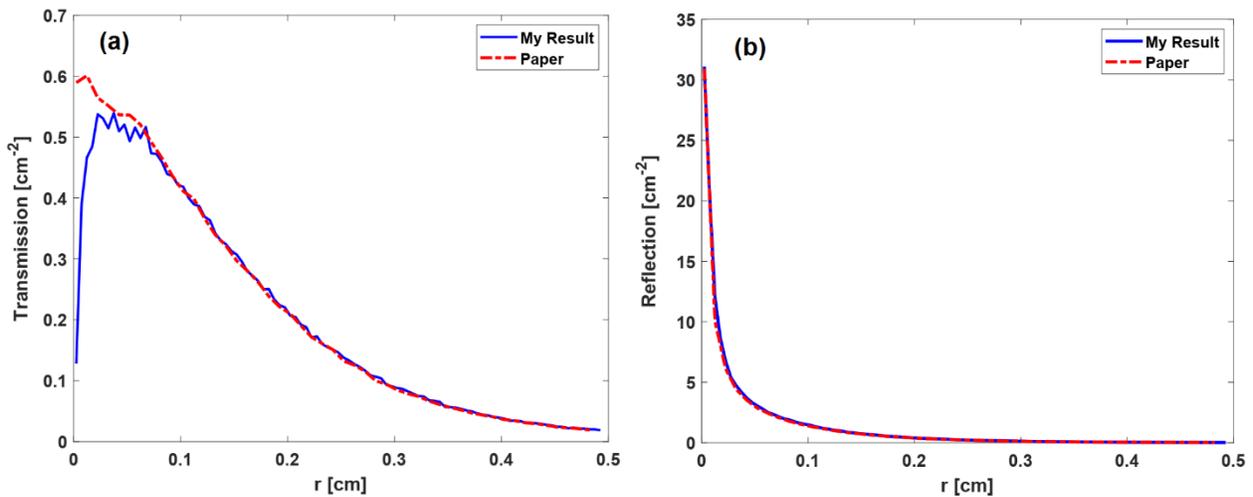

**Fig. 2.10.** Steady-state Monte Carlo simulation of the structure shown in Fig. 2.9. The transmittance (a) and diffuse reflectance (b) as flux densities of escape [1/cm²] versus radial position 'r' at front and rear surfaces, respectively. The blue curves represent the results obtained by our MATLAB program, and the red curves are the results from Ref. [32].

For 2D arrays such as the diffuse reflectance or transmittance as a function of $r$ and $\alpha$, $R_d(r, \alpha)$, $T_t(r, \alpha)$, the outputs are in three columns. The first two columns give the first and the second independent variables, respectively, and the third column gives the physical quantities. The results are shown as the images in Fig. 2.11 (a) and (b) which are recorded at the structure bottom $(z = 0.4 \text{ cm})$ and top $(z = 0)$ surfaces, respectively. The computational time was 4.13 hrs performed on a 64-bit AMD EPYC 7452 8-Core 2.35 GHz central processing unit (CPU). To better illustrate, the trajectory of photons as they propagate through the sample multilayer structure is shown in Fig. 2.12.

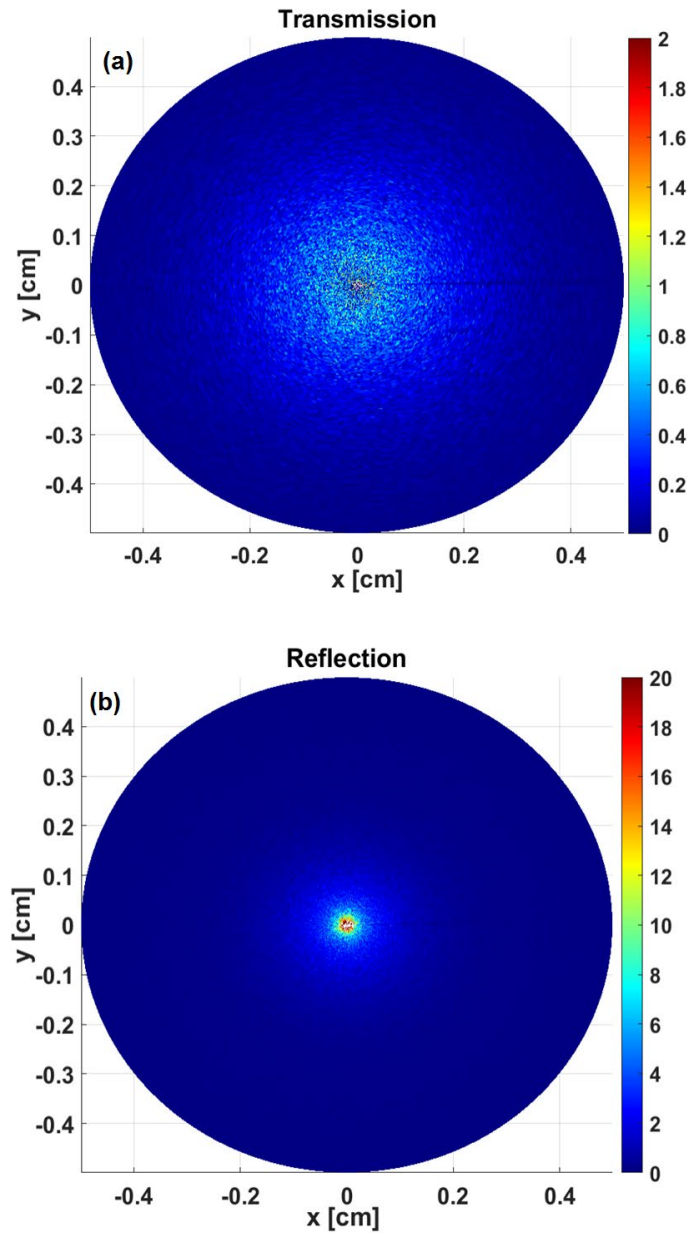

**Fig. 2.11.** The distribution of (a) transmittance and (b) diffuse reflectance on the structure (Fig. 2.9) bottom $(z = 0.4 \text{ cm})$ and top $(z = 0)$ surfaces, respectively.

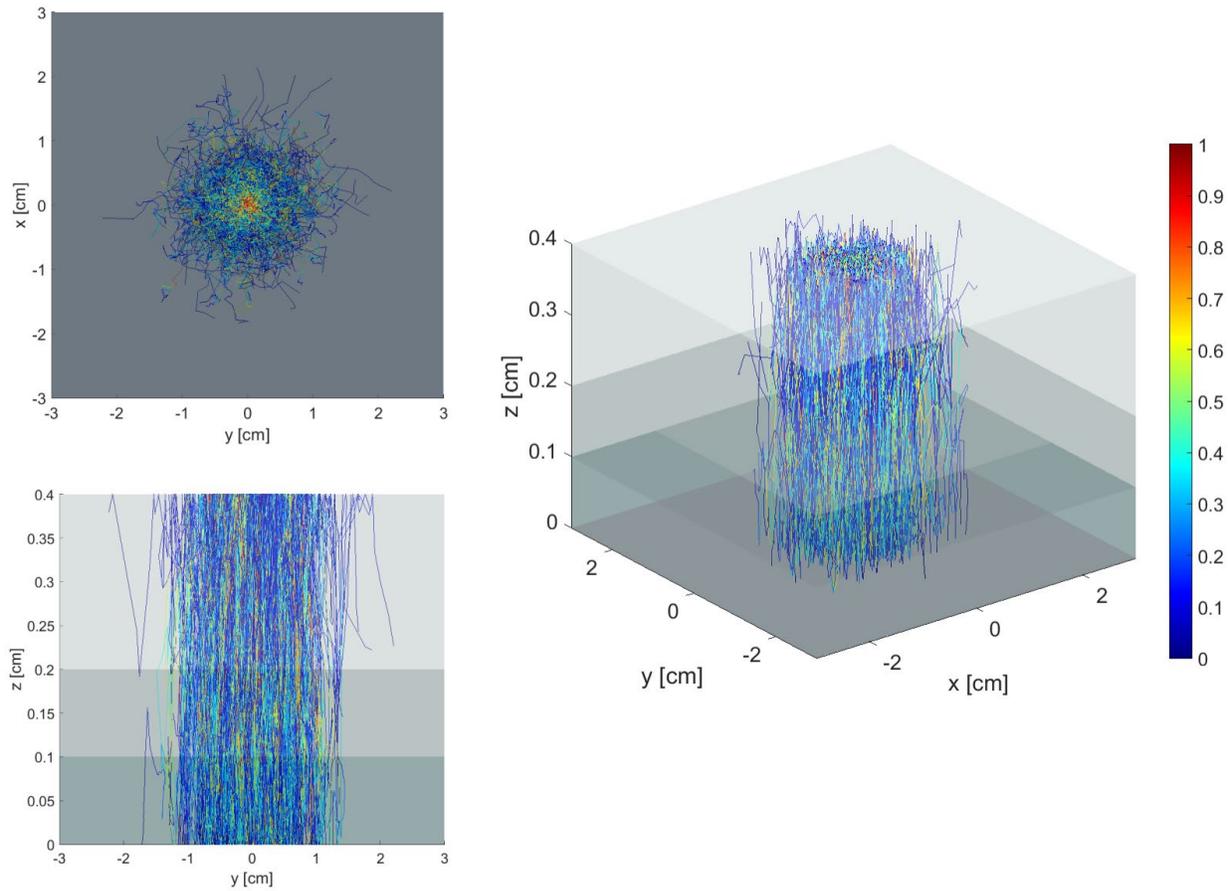

**Fig. 2.12.** Each photon packet trajectory is shown. Color indicates weight of photon packet, where red is the highest and blue is the lowest weight.

To improve the quality of images (reduce the noises observed in 1-D diagrams), one need to increase the number of photons launched to the structure. However, this causes extensive computation and long runtimes, easily taking up to several hours. Initially, the drawback of MCML is the long time taken for each simulation. For more complicated tissue geometries such as tissues with embedded objects, it may take several days or weeks. In addition, when the level of scattering increases, the number of scattering events in each path increases, this increases the time required to simulate the path of one photon.

Generally, when more complicated systems are simulated, the variance in the results increases and more photons have to be tracked to gain a statistically significant result. In recent years, studies on parallel MC algorithms have successfully reduced this computational cost down to seconds or minutes, due largely to the inherent "parallelizable" nature of MC as each photon propagates independently from the others. The following section will discuss this technique in further details.

## 2.6 Parallel Computing (Optimization of the MC Code)

As we mentioned earlier, sequential MC simulations suffer from extensive computational burden. One recent breakthrough in the context of MC algorithm development is to take the advantage of parallel computing capability of modern graphics processing units (GPUs) which were dependent on CUDA programming and dramatically shortens the computational time by 2 to 3 orders of magnitude [23, 33] compared to the highly optimized single-thread CPU implementation. If you don't have a GPU and run the program on MATLAB platform, you can use the capabilities of MATLAB®, Parallel Computing Toolbox™ (PCT) and MATLAB Parallel Server™. This report have done parallel computing based on the MATLAB PCT to reduce the execution time.

Parallel computing allows you to carry out many calculations simultaneously. Large problems can often be split into smaller ones, which are then solved at the same time. The main reason to consider parallel computing is to save time by distributing tasks and executing these simultaneously. With PCT, one can accelerate his code using interactive parallel computing tools, such as *parfor* which is the command we used in our code instead of analogous *for*-loop.

A *parfor*-loop in MATLAB® executes a series of statements in the loop body in parallel. The MATLAB client issues the *parfor* command and coordinates with MATLAB workers to execute the loop iterations in parallel on the workers in a parallel pool. The client sends the necessary data on which *parfor* operates to workers, where most of the computation is executed. The results are sent back to the client and assembled.

Each execution of the body of a *parfor*-loop is an iteration. MATLAB workers evaluate iterations in no particular order and independently of each other. Because each iteration is independent, there is no guarantee that the iterations are synchronized in any way, nor is there any need for this. You cannot use a *parfor*-loop when an iteration in your loop depends on the results of other iterations. Each iteration must be independent of all others. Fig. 2.13 represents the flowchart of performing PCT on MATLAB platform. We refer to Ref. [34] webpage for those interested to understand more details of workflow of this approach.

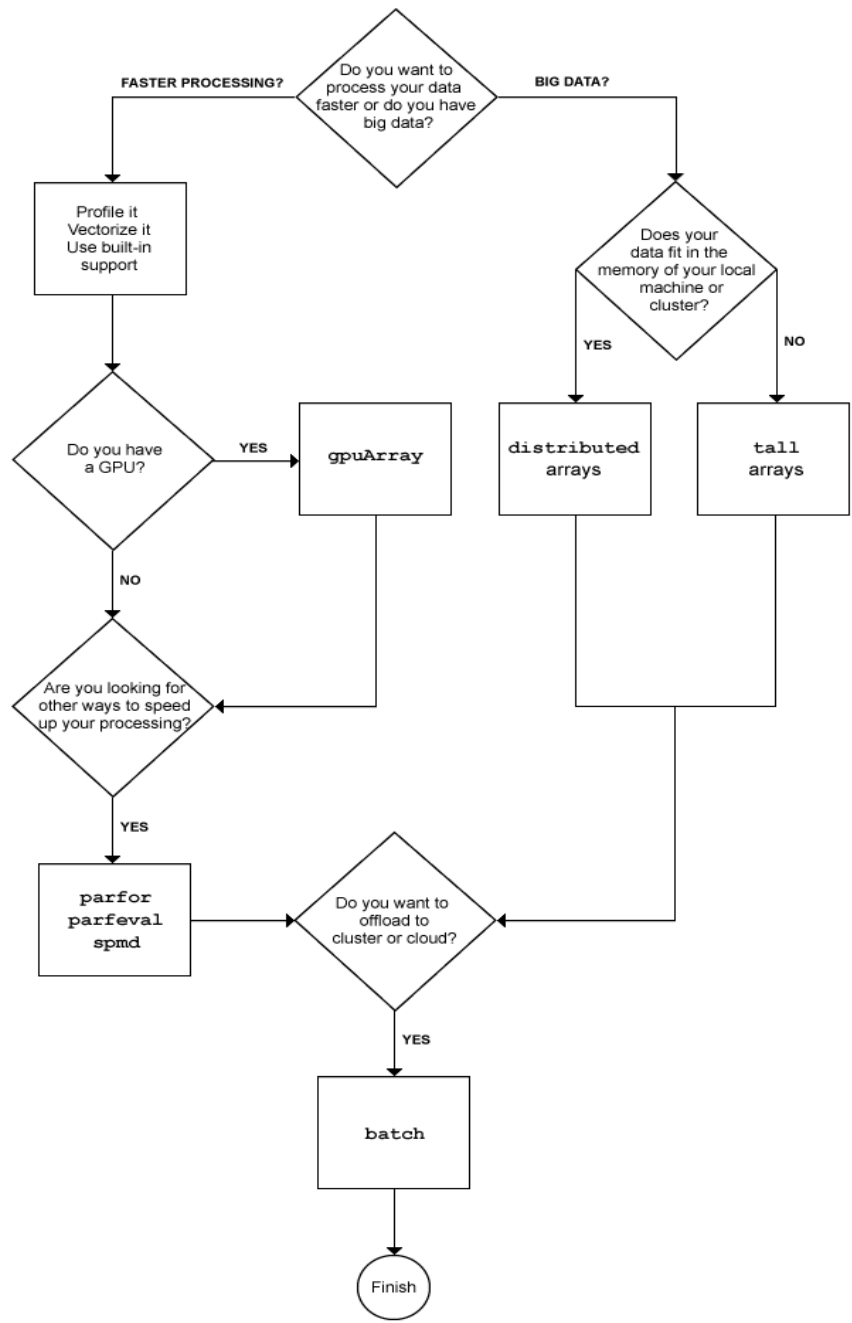

**Fig. 2.13.** The flowchart steps to perform PCT on MATLAB platform

We have modified our previous code which was based on a normal *for*-loop to use *parfor* command instead. The results of these two simulations (*for*- and *parfor*-loop) are shown in Fig. 2.14. It can be seen that both codes yield the same results, but the simulation time for *parfor*-loop running on 12 MATLAB workers was near 28 minutes which is about ×9 times faster than the conventional *for*-loop implementation. This tremendous reduction in computation time make parallel computing an attractive tool in MC simulations of photon migration in tissues. Throughout the rest of this report, all of the simulations are based on the *parfor*-loop execution.

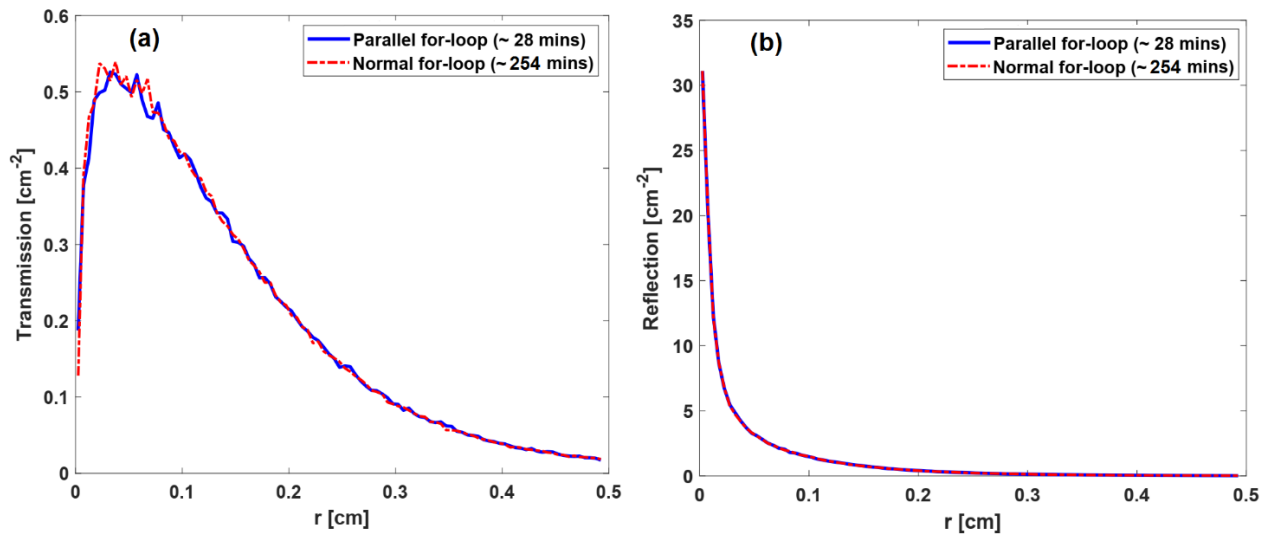

**Fig. 2.14.** Plots of (a) transmittance and (b) diffuse reflectance of the configuration in Fig. 2.9 which are obtained by parallel computing (blue curves) performed using PCT on MATLAB platform. We also include normal *for*-loop outputs for comparison. The simulation times were 28 minutes for the *parfor*-loop and 254 minutes for the normal *for*-loop, i.e., the *parfor* proves to be 9 times faster. Comparing the *parfor*- to normal for-loop results validates the MC code for PCT implementation.

# 3. Sampling of Scattering Phase Function

In MC simulation of photon transport in tissues, a key step is to sample scattering directions from the phase function, which plays a fundamental role in understanding the physical properties of the medium. In recent studies, phase functions have become increasingly important in modeling of the reflectance obtained by spatially-resolved reflectance spectroscopy [35, 36] or spatial-frequency domain reflectance spectroscopy [37-39], in particular for small source-detector separations or high spatial frequencies. Biological tissues have a very complex structure and it is not a trivial matter to decide which phase function will be the best choice. Fortunately, in many cases, if the distance between the source of light and the detector is large enough (e.g. 20–30 mm) [40], the problem simplifies and it becomes possible to reasonably describe the photon scattering by a very general phase function. Hence, the accuracy of the simulated results in MC relies on the sampling of scattering phase functions, and consequently, efficient sampling algorithms are particularly essential and in high demand. This section is organized to evaluate the existing implementations of the scattering phase function sampling for MC simulations.

## 3.1 Phase function

When a photon is scattered, the new propagation direction is calculated with reference to the far-field intensity distribution that a single scatterer would produce from an incoming light beam. The normalized far-field intensity is called the phase function and, within a MC framework, it is assigned to be the probability distribution function of the scattering direction, such that after launching sufficient photons the far-field intensity is reproduced. In general, in a MC simulation, a scattering phase function is defined as an angular probability density function $p\left(\hat{e}_p^i, \hat{e}_p^s\right)$ for a photon deflecting from the propagation (incident) direction $\hat{e}_p^i$ in the direction $\hat{e}_p^s$. Although turbid media are sometimes anisotropic, biological tissues are often assumed to be isotropic, where the scattering depends only on the angle $\theta$ between $\hat{e}_p^i$ and $\hat{e}_p^s$. Consequently, the phase function can be written as $p(\phi, \cos\theta)$, where $\phi$ is the azimuthal angle given that the $z$-axis is aligned with the propagation direction $\hat{e}_p^i$. Under the assumption that $\phi$ and $\cos\theta$ are independent variables, the phase function can be written as $p(\phi,\cos\theta)=p(\phi)p(\cos\theta)$. In the literature, the term phase function is most commonly used for the probability density of $\cos\theta$, since $p(\phi)=1/2\pi$ for symmetric scatterers. Consequently, we will use the term phase function exclusively for $p(\cos\theta)$ throughout the remainder of this chapter.

Many approximate analytic phase functions are widely employed owing to their simplicity and usefulness. The most commonly used phase function in the biomedical community is the Henyey-Greenstein (HG) phase function that allows simple and computationally efficient description of scattering in biological tissues (See Eqs. (2. 29-31)). This definition is computationally fast and simple to interpret. However, since it has been shown that the HG phase function underestimates large-angle scattering [41], other phase functions such as the modified Henyey-Greenstein (MHG) and Gegenbauer kernel (GK) have been proposed for improving the accuracy. The MHG phase function is defined as a weighted sum of the HG phase function and Rayleigh scattering [42] as:

$$p_{MHG}(\cos\theta)=\beta p_{HG}(\cos\theta)+(1-\beta)\frac{3}{2}\cos^2\theta \qquad (3.1)$$

where $\beta$ adjusts their relative contribution. The GK phase function is a mathematical extension of the HG phase function with an additional free parameter $\alpha_{GK}$ [43, 44]:

$$p_{GK}(\cos\theta)=\frac{2\alpha_{GK}g_{GK}\left(1-g_{GK}^2\right)^{2\alpha_{GK}}}{\left[\left(1+g_{GK}^2\right)^{2\alpha_{GK}}-\left(1-g_{GK}^2\right)^{2\alpha_{GK}}\right]\left(1+g_{GK}^2-2g_{GK}\cos\theta\right)^{(1+\alpha_{GK})}} \qquad (3.2)$$

It should be noted that $g_{GK}$ is in general not equal to the anisotropy factor of the GK phase function. The above more complex phase functions are, however, always an approximation of the real phase function.

## 3.2 Analytical sampling of the scattering angles (Inverse CDF)

To sample the $\phi$ and $\cos\theta$ in the MC simulations, the cumulative distribution function (CDF) corresponding to the probability density functions $p(\phi)$ and $p(\cos\theta)$ have to be derived

$$CDF_{\phi}(\phi')=\frac{1}{2\pi}\int_0^{\phi'}d\phi \qquad (3.1)$$

$$CDF_{\cos\theta}(\cos\theta')=\frac{1}{2\pi}\int_{-1}^{\cos\theta'}p(\cos\theta)d\cos\theta \qquad (3.2)$$

where $p(\phi)$ can be set to $1/2\pi$ if all the azimuthal angles are equally likely. The obtained CDFs are then usually equated to uniformly-distributed random numbers $\xi_1$ and $\xi_2$ from [0, 1]. Subsequently, the sampled $\phi$ and $\cos\theta$ can be derived by evaluating the corresponding inverse CDFs as

$$\phi = 2\pi\,\xi_1 \tag{3.3}$$

$$\cos\theta = CDF_{\cos\theta}^{-1}(\xi_2) \tag{3.4}$$

The main point is that the computation of an inverse CDF significantly depends on the form of the phase function. For example, the HG phase function is very convenient since the inverse CDF can be expressed analytically. Consequently, $\cos\theta$ can be sampled from a closed form analytical expression. Analytical forms of the inverse CDF allow simple and computationally efficient implementation in the MC model.

## 3.3 Numerical sampling of the scattering angles (Tabulated method)

For realistic phase functions such as Mie's phase function, an analytic form of their inverse CDF may not exist or computing their values may be too expensive. Therefore, sampling has to be conducted in a different way. Previous approaches employ some form of rejection sampling [45], however, this method is computationally very inefficient in particular for probability distribution containing peaks, which is the case with the forward-peaked phase functions of biological tissues. Alternatively, a lookup table (LUT) based sampling of $\cos\theta$ can be employed. In the following section, we present tabulated methods for accurately sampling random scattering directions in MC radiative transfer.

**3.3.1. Forward lookup table sampling method.** The entries of the forward lookup table (FLT) sampling method are the discrete $CDF_j$ values calculated for each $\cos\theta_j$:

$$CDF_j = \int_{-1}^{\cos\theta_j} p(\cos\theta)\,d\cos\theta \tag{3.5}$$

where $j$ goes from 0 to $M$, $\cos\theta_0 = -1$ and $\cos\theta_M = 1$. Given a random number $\xi$ drawn from a uniform distribution, the task is to find the value of $\cos\theta_\xi$, such that $CDF(\cos\theta_\xi)$ corresponds to $\xi$ (Fig. 3.1(a)). This can be accomplished by finding an index $J$ for which

$$CDF_j < \xi \le CDF_{j+1} \qquad (3.6)$$

Once index $J$ is found, the sampled $\cos\theta_\xi$ can be estimated by a linear interpolation between the $\cos\theta_J$ and $\cos\theta_{J+1}$ values, corresponding to $CDF_J$ and $CDF_{J+1}$:

$$\cos\theta_\xi = \frac{\cos\theta_{J+1} - \cos\theta_J}{CDF_{J+1} - CDF_J}(\xi - CDF_J) + \cos\theta_J \qquad (3.7)$$

The most computationally intensive part of this sampling method is the process of finding index $J$. For a given value of $\xi$ this can be accomplished by one of the well-known root-finding methods such as bisection or Newton-Raphson method. However, searching through a LUT in this way is very time consuming and also depends on the size of the LUT. To avoid the pitfalls of this sampling method, an alternative numerical approach has been presented.

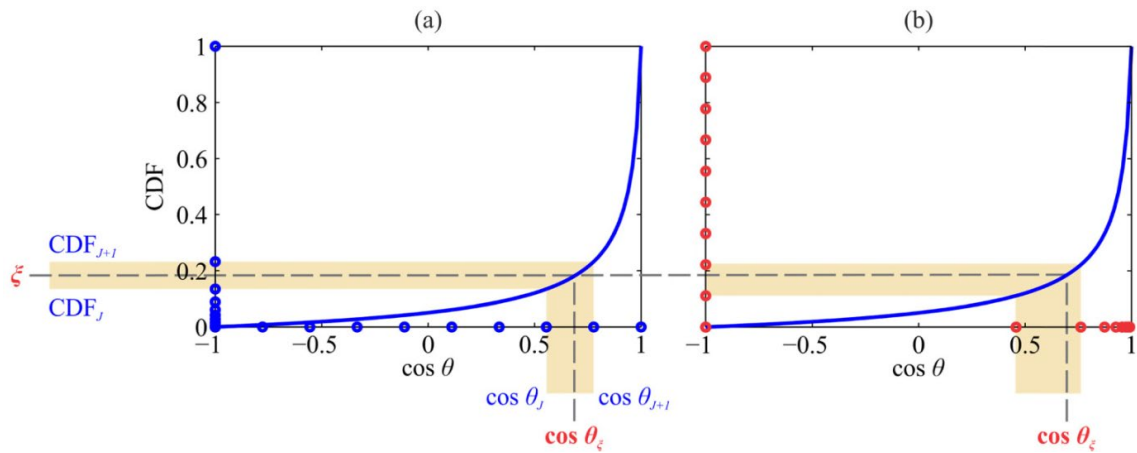

**Fig 3.1.** An example of the CDF obtained for the HG phase function (blue curves). Drawn random number $\xi$ and the sampled $\cos\theta_\xi$ are given in red. (a) The forward lookup table ($\cos\theta_j$ values and their corresponding $CDF_j$ values) is presented with blue circles. The values of $\cos\theta_j$ in the forward lookup table are evenly spaced. (b) The inverted lookup table ($\cos\theta_j$ values and their corresponding $CDF_j$ values) is presented with red circles. The $CDF_j$ values in the inverted lookup table are evenly spaced [46].

**3.3.2. Inverted lookup table sampling method.** In this method, first we obtain a table of CDF values in terms of the evenly-spaced values of $-1 \le \cos\theta \le 1$. The MATLAB program for this purpose is written as:

```
cos_theta = linspace(-1,1,M);              % M: Number of sampling points

p_HG = p_HG(cos_theta);                    % Equation 2.9

p_HG = p_HG/sum(p_HG);                      % normalization

CDF = cumsum(p_HG);                         % cumulative summation

CDF = CDF/max(CDF);                         % normalization

CDF_lin = linspace(0,1,M);

costheta_nl = interp1(CDF,cos_theta,CDF_lin);    % 1-D interpolation

zeta = rand(0,1);

costheta_z = interp1(CDF_lin,costheta_nl,zeta); % linear interpolation of the value of the
                                            % sampled $\cos\theta_\xi$ .
```

This method is an inverted LUT sampling method where the forward lookup table (evenly-spaced $\cos\theta_j$ values and corresponding $CDF_j$ values) is transformed to an inverted lookup table by interpolation to obtain evenly-spaced points of the $CDF_j$ values and corresponding $\cos\theta_j$ values. Subsequently, a linear interpolation can be used to refine the value of the sampled $\cos\theta_\xi$ (Fig. 3.1(b)).

**3.3.3. Sample Computation.** We have used the tabulated method to analyze the configuration shown in Fig. 2.9 again. Sampling from the HG phase function was accomplished with the inverted-LUT sampling method. The number of sampling points here was $M = 2^{16}$. Note that the lookup table size is an important factor affecting the final results. In this particular case, an undersized lookup table leads to an underestimated transmittance and reflectance. On the other hand, constructing a large lookup table for phase functions can be computationally intensive.

The simulation results are shown in Fig. 3.2 and compared with those generated by analytical sampling (section 2) to examine the correctness of our LUT program.

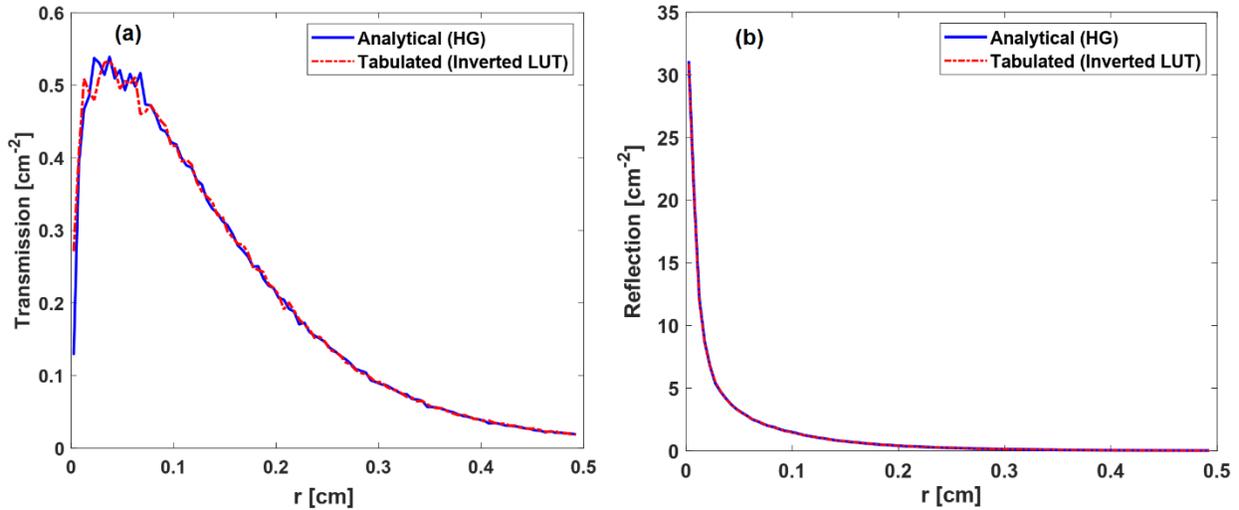

**Fig 3.2.** MCML simulated (a) transmittance and (b) reflectance of the three layer turbid structure (Fig. 2.9) where the phase function was sampled using the inverted-LUT sampling method (red curves). The results are compared with those obtained by analytical sampling method (blue curves) to examine the correctness of our program.

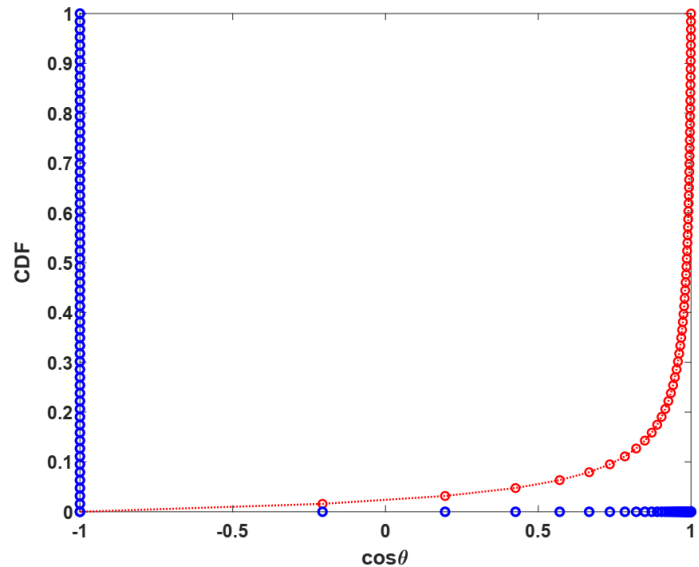

**Fig 3.3.** An example of CDF for the HG phase function (red curve) to clearly illustrate the smapling scheme. The inverted LUT method is used where the values of $CDF_j$ are evenly-spaced and the points of $\cos \theta_j$ have a nonlinear distribution (blue circles). The number of sampling points here is $M = 2^6$.

In Fig. 3.3, our used sampling scheme can be clearly observed in which the blue circles represent the sampling points. Moreover, histogram of the sampled $\cos\theta_{\xi}$ for both methods of analytical and inverted LUT sampling is shown in Fig. 3.4 (a) and (b), respectively.

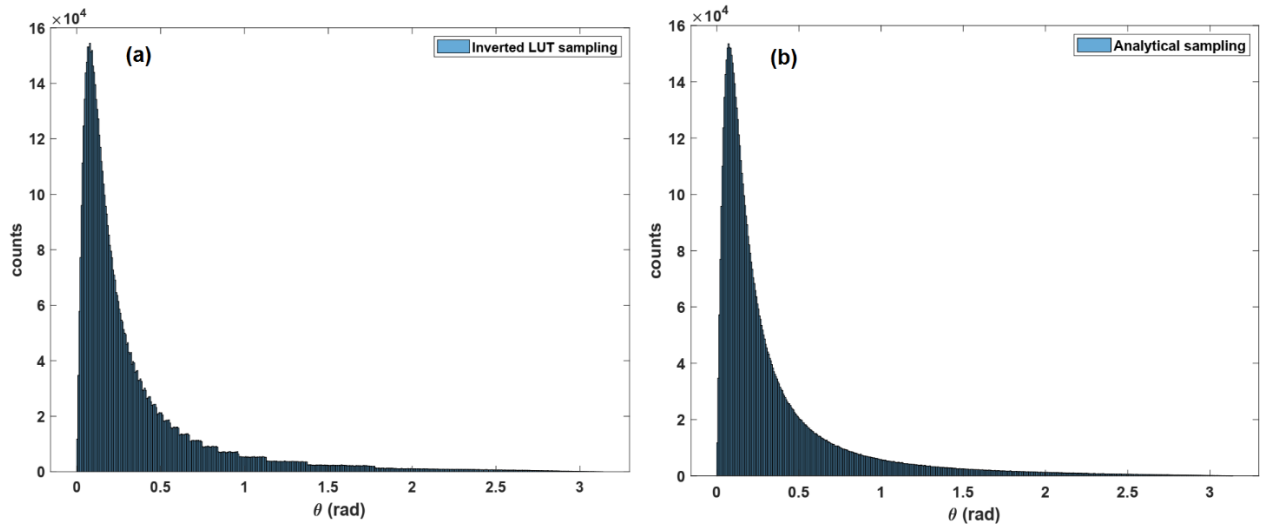

**Fig 3.3.** Histogram of 100000 samples generated using (a) inverted LUT tabulated method, and (b) analytical sampling method. Both of the samplings are based on the HG scattering phase function.

# 4. MCML Simulation with Embedded Objects

Multilayer model was a highly simplified model for many applications. Hence, MCML was further modified to incorporate objects of regular shapes, such as sphere and cylinder. The embedded object was defined by parametric equation. MCML was first modified to handle embedded objects of matched refractive index. Since there were errors in the observed output parameters, MCML was further modified to handle the refractive index mismatch for embedded objects of defined geometry, such as sphere, cylinder, ellipsoid, and cuboid (MCEO). These geometries would be useful in providing a realistic tissue structure such as modelling for lymph nodes, tumors, blood vessels, head and other simulation medium. The steps involved when photons hit the tissue-object boundary where they were either transmitted or reflected based on its refractive indices have been discussed in Ref. [47].

In this section, MCML with object of regular geometry (sphere) was embedded in the layered surrounding tissue; it is referred to this as MMCL with embedded objects (MCML-EO). The refractive index mismatch between the object and the surrounding tissue is also considered in this simulation. To understand the simulation details for more regular geometries, one can refer to Ref. [47].

## 4.1. Mathematical Relations for an Embedded Sphere

For embedded objects there are two challenges: one is computing the distance of the photon from the scattering site to the object boundary and the other is determining the direction cosine of the photon after it is reflected/transmitted at the object boundary. Both the problems arise due to the curved nature of the object. Therefore, there are two things to find out:

- The distance between the photon's current position and the boundary of the object (sphere)
- Check whether the photon is going to hit the boundary with the step-size it needs to travel

To obtain the intersection point between the ray (a packet of photons) and the object, the equation of the ray and the equation of curved object is required. A ray of origin $(O_x, O_y, O_z)$ with direction cosines $d(d_x, d_y, d_z)$ parameterized by distance $t$ is represented by the equation $(\vec{O} + t \cdot \vec{d})$, where "·" implies multiplication of scalar $t$ with every element in vector $\vec{d}$. Any point on the ray at distance $t$ from origin of ray is given by $[(O_x + t * d_x), (O_y + t * d_y), (O_z + t * d_z)]$. Equation of an origin-shifted sphere centered at $(C_x, C_y, C_z)$ with radius $r$ is:

$$(x - C_x)^2 + (y - C_y)^2 + (z - C_z)^2 = r^2 \tag{4.1}$$

The intersection point between the ray and the curve object has to satisfy both of the equation of ray and the equation of curve object. To find this point, the ray equation $(\vec{O} + t \cdot \vec{d})$ should be substituted in the equation of sphere (Eq. (4.1)) where a quadratic relation is obtained as:

$$I \cdot t^2 + J \cdot t + K = 0 \tag{4.2}$$

where $(\cdot)$ is the multiplication of numerical values with variable $t$ to solve the quadratic equation, and,

$$I = d_x^2 + d_y^2 + d_z^2 \tag{4.3}$$

$$J = 2 * (d_x * del_x + d_y * del_y + d_z * del_z) \tag{4.4}$$

$$K = del_x^2 + del_y^2 + del_z^2 - r^2 \tag{4.5}$$

in which $del_x = (O_x - C_x)$, $del_y = (O_y - C_y)$, and $del_z = (O_z - C_z)$. In all the equations (*) denotes the multiplication of two numbers. The ray would intersect with a curved object, if the solution to the quadratic equation in $t$ [Eq. (4.2)] is real, i.e., $(J^2 - 4 * I * K) > 0$. If, $(J^2 - 4 * I * K) < 0$, then the ray does not intersect the curved object. In that case, there is no hit. Since there are two solutions to the quadratic equation, there are two points of intersection. The distance between the points of intersection $(\vec{O} + t_{int} \cdot \vec{d})$ and origin of ray $\vec{O}$ is computed. If the distance is greater than the step-size, there is a boundary hit. So, the algorithm is as follow:

$$\begin{cases} \text{if} \quad t_{int} > \text{photon.step\_size} \;\Rightarrow\; \text{No Hit} \\ \text{if} \quad t_{int} < \text{photon.step\_size} \;\Rightarrow\; \text{Hit Boundary} \;\Rightarrow\; \text{Obtain distace to boundary} \;\Rightarrow\; \Delta\text{s}' = \Delta\text{s}_{to-boundary} \end{cases}$$

In case of a boundary hit, the photon position coincides with the intersection point by taking the step of $\Delta s'$. Once the photon reaches the boundary, it will undergo either reflection or transmission. For layer interface, the normal to tangent plane coincides with the global z axis. However, in case of curved geometry, the normal to tangent plane does not coincide with the global coordinate $z$-axis. Therefore, it needs to be transformed into a local coordinate system whose $z$-axis matches with the normal to tangent. Hence, reflection/transmission is done in local coordinate system and then converted back to the global coordinate just as it is done for spinning of photon in MCML. A local Cartesian coordinate system is created with the normal line (line perpendicular to the tangential plane) as $z$ axis. The steps involved in conversion of global coordinate system $(\mu_x, \mu_y, \mu_z)$ to local system $(\mu'_x, \mu'_y, \mu'_z)$ are as follows:

- Rotate $(\mu_x, \mu_y, \mu_z)$ by an angle $\phi_0$ in the positive sense (right hand rule) about positive z-axis to get intermediate coordinates $(\mu''_x, \mu''_y, \mu''_z)$. By inspection,

$$[\mu''] = [R_{\phi_0}][\mu] \Rightarrow \begin{pmatrix} \mu''_x \\ \mu''_y \\ \mu''_z \end{pmatrix} = \begin{pmatrix} \cos\phi_0 & -\sin\phi_0 & 0 \\ \sin\phi_0 & \cos\phi_0 & 0 \\ 0 & 0 & 1 \end{pmatrix} \begin{pmatrix} \mu_x \\ \mu_y \\ \mu_z \end{pmatrix} \quad (4.6)$$

- Rotate $(\mu''_x, \mu''_y, \mu''_z)$ by an angle $\theta$ about the $\mu''_y$ axis to get $(\mu'_x, \mu'_y, \mu'_z)$. This gives:

$$[\mu'] = [R_{\theta_0}][\mu''] \Rightarrow \begin{pmatrix} \mu'_x \\ \mu'_y \\ \mu'_z \end{pmatrix} = \begin{pmatrix} \cos\theta_0 & 0 & \sin\theta_0 \\ 0 & 1 & 0 \\ -\sin\theta_0 & 0 & \cos\theta_0 \end{pmatrix} \begin{pmatrix} \mu''_x \\ \mu''_y \\ \mu''_z \end{pmatrix} \quad (4.7)$$

Combining the rotations by matrix multiplication gives the coordinates of $[\mu']$ with respect to the rotated system $[\mu]$ as:

$$[\mu'] = [R_{\theta_0}][R_{\phi_0}][\mu] \Rightarrow \begin{pmatrix} \mu'_x \\ \mu'_y \\ \mu'_z \end{pmatrix} = \begin{pmatrix} \cos\theta_0\cos\phi_0 & -\sin\phi_0 & \sin\theta_0\cos\phi_0 \\ \cos\theta_0\sin\phi_0 & \cos\phi_0 & \sin\theta_0\sin\phi_0 \\ -\sin\theta_0 & 0 & \cos\theta_0 \end{pmatrix} \begin{pmatrix} \mu_x \\ \mu_y \\ \mu_z \end{pmatrix} \quad (4.8)$$

We define the rotation matrix as:

$$R(\theta_0,\phi_0) = \begin{pmatrix} \cos\theta_0\cos\phi_0 & -\sin\phi_0 & \sin\theta_0\cos\phi_0 \\ \cos\theta_0\sin\phi_0 & \cos\phi_0 & \sin\theta_0\sin\phi_0 \\ -\sin\theta_0 & 0 & \cos\theta_0 \end{pmatrix} \tag{4.9}$$

The polar and azimuthal rotation angles $(\theta_0,\phi_0)$ can be assessed from:

$$\cos\theta_0 = \frac{z_i - C_z}{r}, \quad \sin\theta_0\cos\phi_0 = \frac{x_i - C_x}{r} \tag{4.10}$$

where $r$ is the sphere radius, and $(x_i,y_i,z_i)$ and $(C_x,C_y,C_z)$ mark the coordinates of the photon incidence point (intersection point) and sphere center, respectively, in the global coordinate system.

Direction vector of the incident photon, $[\mu]$, is thus expressed in the rotated coordinate system as

$$[\mu'] = [R(\theta_0,\phi_0)][\mu] \tag{4.11}$$

The direction vector after reflection from the boundary, $\left[(\mu_R)'\right]$, is now obtained simply by changing the sign of component $\mu'_z \rightarrow -\mu'_z$. Upon transformation back to the global coordinate system (by multiplying with the inverse of matrix R), we obtain:

$$[\mu_R] = [R(\theta_0,\phi_0)]^{-1}\left[(\mu_R)'\right] = [R(\theta_0,\phi_0)]^{-1}\begin{pmatrix} \mu'_x \\ \mu'_y \\ -\mu'_z \end{pmatrix} \tag{4.12}$$

Note that the inverse of the rotation matrix, $[R(\theta_0,\phi_0)]^{-1}$ is its transpose, $[R(\theta_0,\phi_0)]^T$, that is:

$$[R(\theta_0,\phi_0)]^{-1} = [R(\theta_0,\phi_0)]^T \tag{4.13}$$

Or

$$R_{ij}(\theta_0,\phi_0)^{-1} = R_{ji}(\theta_0,\phi_0) \tag{4.14}$$

The normal to the tangential plane is needed for Fresnel computations. Angle of incidence $\alpha_i$ is $\cos^{-1} |\mu_z|$. When there is a refractive index match, angle of transmittance $\alpha_t$ is equal to $\alpha_i$. In case of refractive index mismatch, Snell's law is used to compute $\alpha_t$ as:

$$n_i \sin \alpha_i = n_t \sin \alpha_t \tag{4.15}$$

where $n_i$ and $n_t$ are the refractive indices of the medium of incidence and medium of transmittance. If $n_i > n_t$ and $\alpha_i$ greater than the critical angle $\alpha_{cr} = \sin^{-1}(n_i/n_t)$, probability of reflection $R(\alpha_i)$ is unity. Otherwise, Fresnel's formula [Eq. (2.19)] for un-polarized light determines the percentage of photon being reflected and the rest is transmitted.

During transmission, the direction cosines of the photon are updated in the rotated coordinate system as follows:

$$\left( \mu_x^T \right)' = \frac{n_1}{n_2} \mu_x'$$

$$\left( \mu_y^T \right)' = \frac{n_1}{n_2} \mu_y'$$

$$\left( \mu_z^T \right)' = \cos \alpha_t \, \text{sgn}(\mu_z') = \sqrt{1 - \sin^2 \alpha_t} \, \text{sgn}(\mu_z')$$

$$= \sqrt{1 - \left( \frac{n_1}{n_2} \right)^2 \left[ 1 - \left( \mu_z' \right)^2 \right]} \, \text{sgn}(\mu_z')$$

<div align="right">(4.16)</div>

where $\text{sgn}(\mu_z')$ marks the sign function. The refracted direction vector expressed in the global coordinate system is then obtained analogously to the above:

$$\left[ \mu_T \right] = \left[ R(\theta_0, \phi_0) \right]^{-1} \left[ \left( \mu_T \right)' \right] \tag{4.17}$$

Figure 4.1 depicts the flow chart of MCML modified for embedded object to better illustrate all of the steps involved.

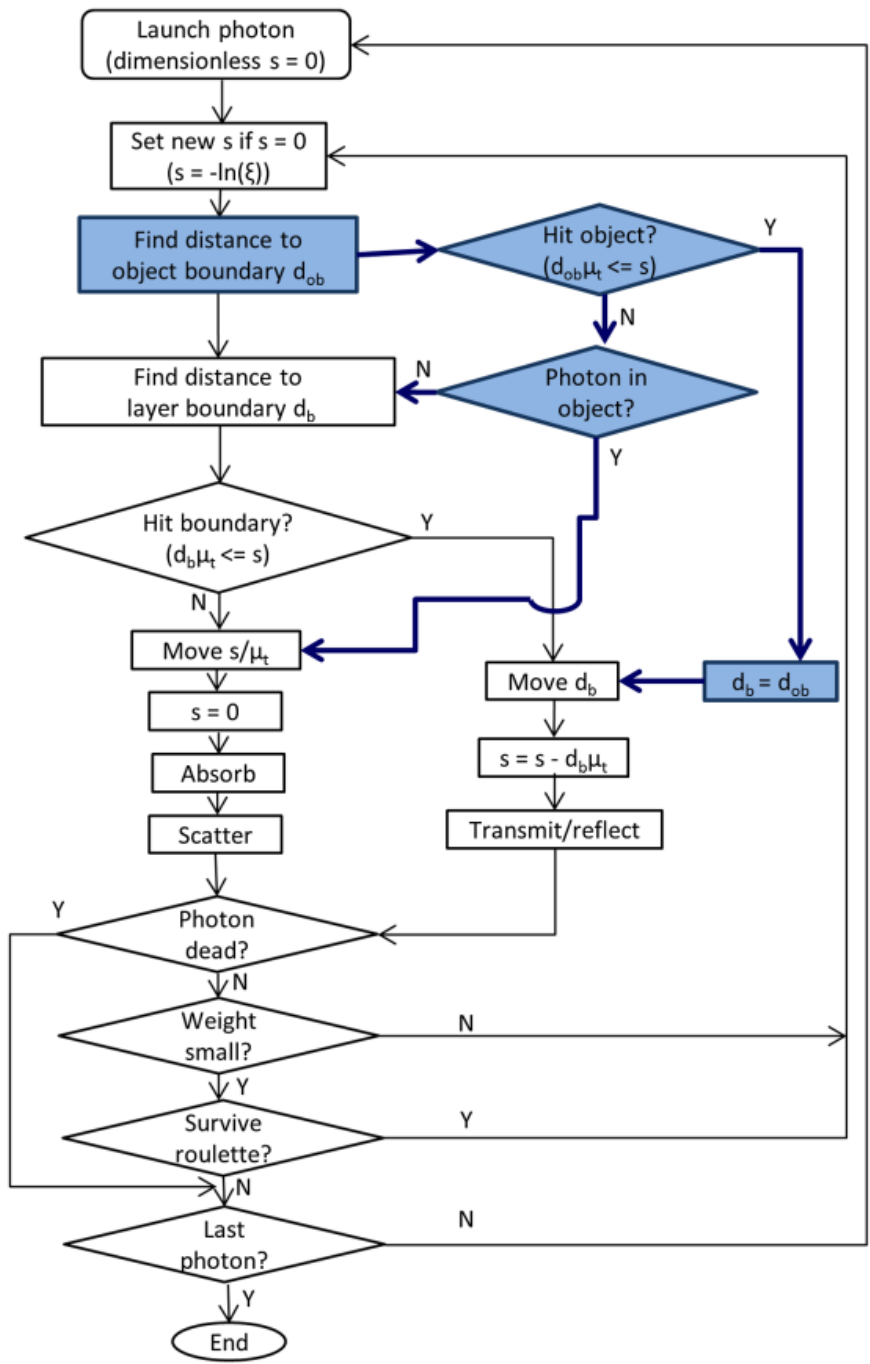

**Fig 4.1.** A flow chart depicting involved steps in MC simulation of photon propagation in multilayer tissue with embedded objects.

## 4.2. Sample Computation

The MCML-EO sample computed here is for an embedded sphere of radius $r = 0.5$ cm and centrally located at a depth of 1.2 cm. The dimensions and the optical properties for the embedded object and the surrounding tissue are given in Table 1. For both of the layer and embedded object, $g = 0.9$. The outer medium (launch surface and transmit surface) was taken as air with refractive index 1.0. The depth of the surrounding tissue layer was taken as 6 cm. Fig 4.2 shows the simulation geometry. A pencil beam of light was launched at the origin $(0,0,0)$ along the z-axis with direction cosines $(0,0,1)$. Weight dropped was accumulated in $x$, $y$, and $z$ grids. Number of grids in each axis is 301 and the size of the grid was 0.02 cm. The volume spanned is $-3$ to $+3$ cm in $x$ and $y$ axis. Grid covers 0 to 6 cm in $z$ axis. To display absorbance map $y = 0$ plane is presented. All simulations were run for $10^6$ photons and on the MATLAB interface where a parallel computation was applied.

| Embedded object | Depth (cm) | Dimension required as input parameter | Dimension (cm) | Contains |
|---|---|---|---|---|
| Sphere | 1.2 | Radius | 0.5 | Methylene blue |

| Optical properties of medium | Refractive index ($n$) | Absorption coefficient of medium $\mu_a$ (cm$^{-1}$) | Scattering coefficient of medium $\mu_s$ (cm$^{-1}$) | Scattering anisotropy ($g$) |
|---|---|---|---|---|
| Surrounding tissue | 1.4 | 0.2525 | 254 | 0.9 |
| Methylene blue | 1.3 | 1.7049 | 180 | 0.9 |

**Table 1.** Dimensions, location, and optical properties of the embedded sphere as well as various layers used in the MCML-EO simulation model at 664-nm wavelength.

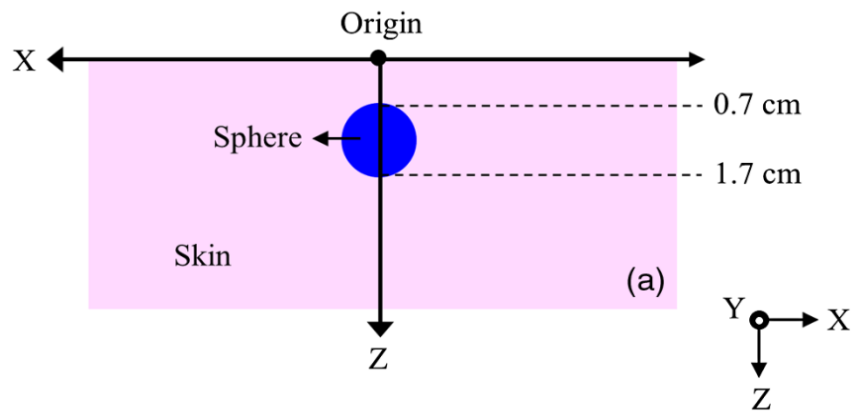

**Fig. 4.2.** Schematic representation of the simulation geometry

Fig. 4.3 shows the absorbance map (in log scale) of our simulation along $y = 0$. The dotted black lines represent the boundary of the embedded object. The structure was simulated using PCT on the MATLAB platform (section 2.6) and the runtime was about 6.5 hrs, performed on a desktop with 64-bit AMD EPYC 7452 8-Core processor and 12 workers. Absorption map is important to study effects of illumination during phototherapy. One can play with the light delivery configuration, optical properties, and various object sizes to see what kind of absorption map one wants to achieve. Moreover, the calculated normalized images of diffuse reflectance $R_d(x, y)$ $[1/cm^2]$ (in log scale) is shown in Fig. 4.4. The transmittance $T_t(x, y)$ $[1/cm^2]$ is not shown here, because the depth of the tissue is very large and photons do not reach the bottom surface of the tissue.

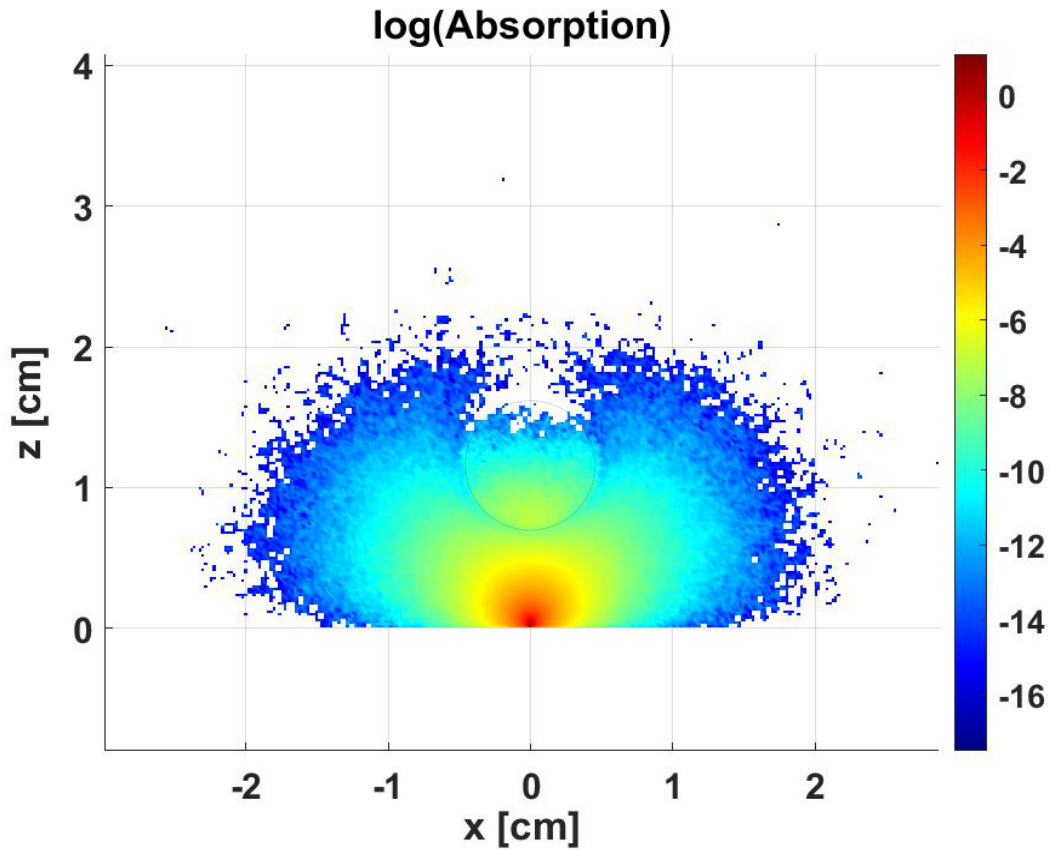

**Fig. 4.3.** Absorption map from MCML for turbid medium with embedded sphere along $y = 0$ plane. Black dotted lines are boundary of the embedded object

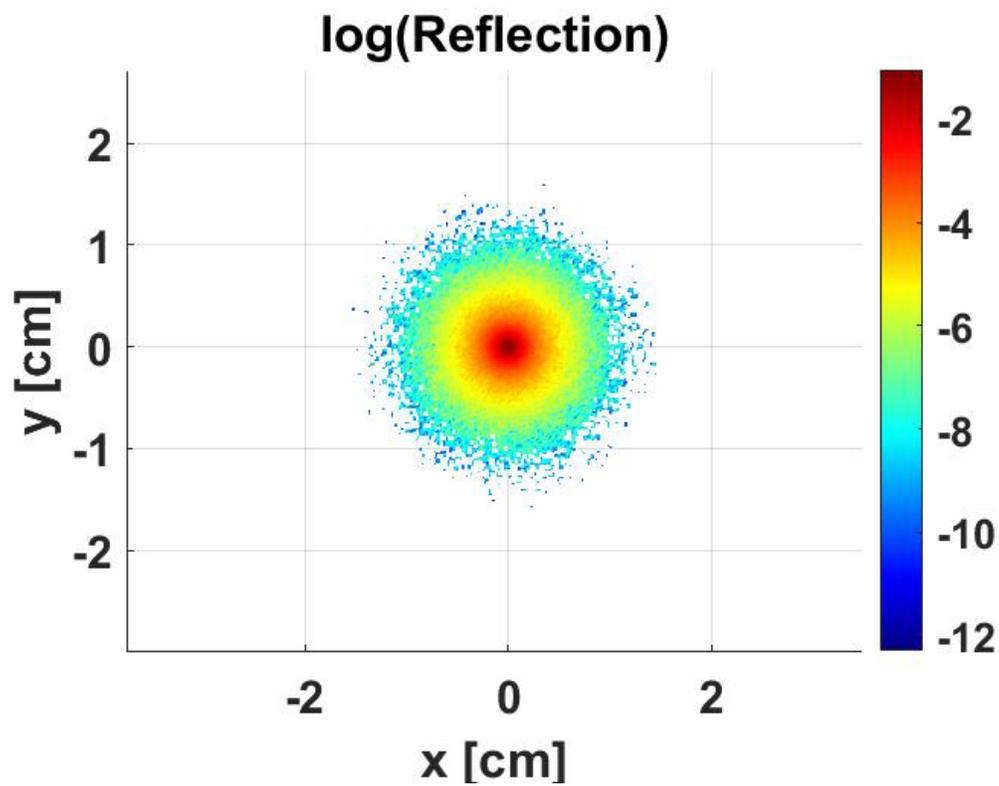

**Fig. 4.4.** Distribution of the simulated MCML-EO diffuse reflectance for the photons detected at the top surface of the tissue ($z = 0$).

# 5. Conclusion and Outlook

The Monte Carlo modelling produces accurate results when comparing transmission and reflection to results obtained from other implementations of MC photon transport software. This report has attempted to provide a thorough study of MCML simulation of light propagation in turbid media. Section 2 has described the rules of steady-state light transport in multi-layered turbid media. We have verified some of the MCML computational results with those of other investigators. In addition, we have reduced computational time from several hours to minutes by using parallel computing toolbox (PCT) provided in MATLAB. This feature is largely due to the parallelizable nature of MC simulation. In section 3, we have evaluated the existing implementations of the numerical phase function sampling for MC simulations and made use of an improved sampling scheme utilizing linearly spaced tabulated CDF values. We have shown the accuracy of our inverted LUT MC program by comparing the results of simulated transmittance and reflectance with those obtained from analytical phase function sampling. Finally, MCML tissue modeling has been modified in section 4 to incorporate object of sphere shape with a refractive-index mismatched boundary. If the embedded object (inhomogeneity) is of regular geometry (shape), then MCML-EO is a good option in terms of computational burden.

This report presents a review of fundamentals of MCML simulation. Over the recent years, several improvements and investigations of MCML has been presented by researchers. In terms of the illumination configuration, one can investigate the effects of the shape, size, polarization, and incident angle of the illumination beam and the detection angle on the diffuse reflectance. In terms of the tissue structure, modeling of tissues nonlinearity and fluorescence with MC algorithm can be explored. In terms of the computational cost, there has been a number of attempts to accelerate simulation time using parallel computing capability of modern graphics processing units (GPUs). As a future work, one can conduct a study in any of these areas and make an improvement.

# Acknowledgments

I would like to express my sincere gratitude to my advisors, Prof. Michael Somekh and Prof. Amanda Wright, for their valuable guidance, support, and mentorship throughout the course of this work. Their profound knowledge, insightful feedback, and unwavering encouragement have been instrumental in shaping this report.

I am deeply grateful for the opportunities they have provided me and for their confidence in my abilities. Their mentorship has been invaluable in my academic and professional growth, and I am honored to have had the privilege of working under their tutelage.